\documentclass[12pt,preprint,useAMS]{aastex}
\usepackage{graphicx}
\usepackage{natbib,graphicx,amssymb}
\slugcomment{draft}

\bibpunct[, ]{(}{)}{;}{a}{}{,}
\def \aj {AJ}
\def \mnras {MNRAS}
\def \pasp {PASP}
\def \apj {ApJ}
\def \apjs {ApJS}
\def \apjl {ApJL}
\def \aap {A\&A}
\def \nat {Nature}
\def \araa {ARAA}
\def \iaucirc {IAUC}
\newcommand{\kms} {$\mathrm{km\;s^{-1}}\,$}

\def\lesssim{\mathrel{\hbox{\rlap{\hbox{\lower4pt\hbox{$\sim$}}}\hbox{$<$}}}}
\def\gtrsim{\mathrel{\hbox{\rlap{\hbox{\lower4pt\hbox{$\sim$}}}\hbox{$>$}}}}
\newcommand{\halpha} {$\mathrm{H\alpha}$}
\newcommand{\ang} {$\mathrm{\AA}\,$}

\begin{document}
\received{}
\revised{}
\accepted{}
\ccc{}
\cpright{}

\title{Spectropolarimetry of the Type IIb Supernova 2001ig
\footnote{Based on observations made with ESO Telescopes at the Paranal Observatory, under programmes 68.D-0571 and 69.D-0438.}}
\shorttitle{Spectropolarimetry of SN 2001ig}
\shortauthors{MAUND et al.}
\author{Justyn R. ~Maund\altaffilmark{1},
J. Craig Wheeler\altaffilmark{1},
Ferdinando Patat\altaffilmark{2}, 
Lifan Wang\altaffilmark{3},
Dietrich Baade\altaffilmark{2} and
Peter A. H\"{o}flich\altaffilmark{4}}
\altaffiltext{1}{Department of Astronomy and McDonald Observatory, The University of Texas, 1 University Station C1402, Austin, Texas 78712-0259, U.S.A.; jrm$@$astro.as.utexas.edu; wheel$@$astro.as.utexas.edu}
\altaffiltext{2}{ESO - European Organisation for Astronomical Research in the Southern Hemisphere, Karl-Schwarzschild-Str.\ 2, 85748 Garching b.\ M\"unchen, Germany; fpatat$@$eso.org; dbaade$@$eso.org}
\altaffiltext{3}{Department of Physics, Texas A\&M University, College Station, Texas 77843-4242, U.S.A.; wang$@$physics.tamu.edu}
\altaffiltext{4}{Department of Physics, Florida State University, Tallahassee, Florida 32306-4350, U.S.A.; pah$@$astro.physics.fsu.edu}

\begin{abstract}
We present spectropolarimetric observations of the Type IIb SN 2001ig
in NGC 7424; conducted with the ESO VLT FORS1 on 2001 Dec 16, 2002 Jan
3 and 2002 Aug 16 or 13, 31 and 256 days post-explosion.  These
observations are at three different stages of the SN evolution: (1)
The hydrogen-rich photospheric phase, (2) the Type II to Type Ib
transitional phase and (3) the nebular phase.  At each of these
stages, the observations show remarkably different polarization
properties as a function of wavelength.  We show that the degree of
interstellar polarization is 0.17\%.  The low intrinsic polarization
($\sim0.2\%$) at the first epoch is consistent with an almost
spherical ($<10\%$ deviation from spherical symmetry) hydrogen
dominated ejecta.  Similar to SN 1987A and to Type IIP SNe, a sharp
increase in the degree of the polarization ($\sim1\%$) is observed
when the outer hydrogen layer becomes optically thin by day 31; only
at this epoch is the polarization well described by a ``dominant
axis.''  The polarization angle of the data shows a rotation through
$\sim40\degr$ between the first and second epochs, indicating that the
asymmetries of the first epoch were not directly coupled with those
observed at the second epoch.  For the most polarized lines, we
observe wavelength-dependent loop structures in addition to the
dominant axis on the $Q-U$ plane.  We show that the polarization
properties of Type IIb SNe are roughly similar to one another, but
with significant differences arising due to line blending effects
especially with the high velocities observed for SN 2001ig.  This
suggests that the geometry of SN 2001ig is related to SN 1993J and
that these events may have arisen from a similar binary progenitor
system.
\end{abstract}

\keywords{supernovae: general --- supernovae: individual: (SN 2001ig) --- techniques:polarimetric}

\section{Introduction}
\label{intro}
Asymmetries are thought to be inherent with core-collapse Supernovae
(CCSNe) and are inferred through observations of runaway O stars and
pulsars \citep{2005MNRAS.364...59D}, spectral line profiles
\citep{2005Sci...308.1284M} and, in the extreme, through the link
between CCSNe and highly-collimated jets that give rise to Gamma Ray
Bursts (GRBs; \citealt{2006ARA&A..44..507W}).  Spectropolarimetry has
been used to directly measure the geometries of both CCSNe
\citep{2003ApJ...592..457W,2006Natur.440..505L} and Type Ia SNe
\citep{2003ApJ...591.1110W}.  While \citet{2005Sci...308.1284M}
provide an example of studying asymmetries in CCSNe by examining the
shape of line profiles in ordinary spectroscopic observations,
spectropolarimetric observations can provide a straightforward and
unambiguous measure of asymmetries \citep{1982ApJ...263..902S}.  The
former technique can only be conducted at late times during the
nebular phase, when the ejecta is optically thin and the spectrum is
dominated by broad emission lines of fordidden transitions.  At this
later epoch the profiles of these lines can trace the geometry of the
ejecta, but are dominated by the Circumstellar Medium (CSM)-ejecta
interaction and are no longer necessarily an accurate tracer of the
geometry of the explosion mechanism.  In addition, during the early,
optically-thick phases the profiles of lines are only weakly dependent
on the geometry \citep{2001ApJ...556..302H}.  Alternatively
spectropolarimetry can be conducted at any epoch, especially at early
times when the geometry of the ejecta is dictated by the shape of the
progenitor and the explosion mechanism.\\ For a spherically-symmetric,
electron scattering atmosphere, the polarization vectors of the light
received, from different portions of the SN, cancel out leading to no
net observed polarization.  For geometries that depart from the
spherical symmetry, there is incomplete cancellation of the
polarization vectors which leads to a net observed linear intrinsic
polarization \citep{1982ApJ...263..902S,
1984MNRAS.210..829M,1984MNRAS.210..839M}.  Electron scattering is
wavelength independent and the measured level of continuum
polarization is associated with the asymmetry of the photosphere.  In
addition, the polarization associated with spectral lines, through
line scattering, yields information concerning the geometry of the
line forming region, and specifically the distribution of particular
elements within the SN ejecta \citep{2003ApJ...592..457W}. \\ SNe are
principally classified by the absence (Type I) or presence (Type II)
of hydrogen in their early time spectra.  All types of SNe except one
(Type Ia) are associated with the core-collapse induced explosion at
the ends of the lives of massive stars.  \citet{2001ApJ...550.1030W}
showed that, in general, Type II SNe are polarized at much higher
levels compared with Type Ia SNe; at $\sim1\%$ and $\sim0.2-0.3\%$
respectively.  Spectropolarimetric studies of the peculiar Type II SN
1987A \citep{1991ApJS...77..405J} and the Type IIP SN 1999em
\citep{2002Msngr.109...47W} showed a significant increase, from 0.1\%
to 1\%, in the level of the polarization as the time since explosion
increased, suggesting increasing asymmetries at increasing depths into
the SNe.  In the case of the Type IIP SN 2004dj
\citep{2006Natur.440..505L} the polarization properties were directly
correlated to the evolution of the SN light curve.  During the
optically thick phase, the polarization remained at a nearly constant
level; increasing dramatically when the SN entered the optically thin
stage, when the complicated internal structure of the core was
revealed.  Type Ibc SNe demonstrate significantly higher levels of
polarization.  Spectropolarimetric observations of the Type Ic SN
2002ap showed intrinsic continuum polarization levels $>1\%$, which
\citet{2002PASP..114.1333L} suggest arises from an asymmetry of at
least 20\%.  In addition, the differences in polarization angles
between features due to Fe and other elements demonstrated the
different distributions of these elements within the ejecta.
\citet{2003ApJ...592..457W} compared the O I 7774$\mathrm{\AA}$ line
with Fe II lines to show that these elements had very different
geometries, and \citet{2002PASP..114.1333L} similarly studied the
difference between Ca lines and those of Fe group elements.\\ The case
of spectropolarimetric observations of Type IIb SNe is particularly
interesting, as this type of SN is the link relating H-rich and
H-deficient CCSNe.  At early times Type IIb SNe show hydrogen in their
spectra, but these features weaken over a period of weeks, leading to a
hydrogen deficient Type Ib SN \citep{1993ApJ...415L.103F,
1993Natur.365..232S}.  The spectroscopic evolution of Type IIb SNe is
due to the presence of a thin veil of hydrogen surrounding the core of
the progenitor, which is observed at early times but is subsequently
dispersed.  The principal example of spectropolarimetry of a Type IIb
SN comes from observations of SN 1993J.  \citet{1993ApJ...414L..21T}
and \citet{1997PASP..109..489T} observed polarizations of $\gtrsim
1\%$, requiring major to minor axis ratios of the emitting region to
be between $1.54$ and $2$. \citet{1995ApJ...440..821H} modelled these
polarizations as being due to asymmetries of the outer H-rich envelope
and an off-centre energy source, and \citet{1997PASP..109..489T}
presented the case of an asymmetric circumstellar medium.  This is
particularly important for the understanding of how the core-collapse
mechanism can give rise to a range of SN types.
\citet{1996ApJ...462L..27W} also showed how, in the case of SN 1987A,
the technique of spectropolarimetry may be used to probe the nature of
the dust in the immediate vicinity of SNe at late times.\\ The
observations of the Type IIb SN 2001ig and the reduction of the data
are given in Section \ref{obs}.  The results of these observations are
presented in \S\ref{results}, and are discussed in \S\ref{disc} in the
context of SN 2001ig as a CCSN and in the broader context of previous
spectropolarimetric observations of other CCSNe, in particular of Type
IIb SNe.

\section{Observations and Data Reduction}
\label{obs}
\begin{figure}
\rotatebox{-90}{
\includegraphics[width=10cm]{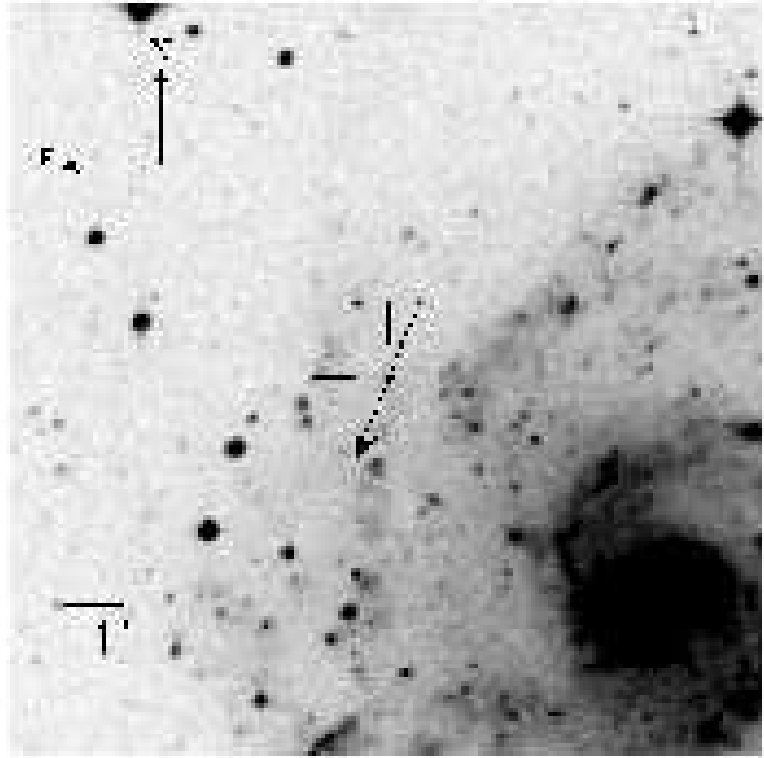}}
\caption{Digital Sky Survey image of the location of SN 2001ig, shown by the filled black circle and cross hairs, relative to its host galaxy NGC 7424. The arrow shows the orientation of the Interstellar Polarization component determined in Section \ref{isp}.}
\label{fig01igpos}
\end{figure}

SN 2001ig was discovered by \citet{2001IAUC.7772....1E} on 2001 Dec
10.43UT, in the galaxy of NGC 7424.  SN 2001ig is located at
$\mathrm{\alpha_{2000} =22^{h}57^{m}30\fs 7}$ and
$\mathrm{\delta_{2000}=-41\arcdeg02\arcmin 26\arcsec}$
\citep{2001IAUC.7777....2R}, which corresponds to an offset of
$\mathrm{185\arcsec E}$ and $\mathrm{108\arcsec N}$ from the centre of
the host galaxy.  The position of SN 2001ig, relative to its host
galaxy, is shown as Fig. \ref{fig01igpos}.
\citet{2004MNRAS.349.1093R} estimated an explosion date of 2001 Dec 3
(JD452247.5), from models of the radio light curve; we adopt this as
the reference date for our observations.  \citet{2001IAUC.7772....2P}
provisionally identified SN 2001ig as being of Type IIb, resembling SN
1987K.  Furthermore, \citet{2002IAUC.7793....2C} observed the
spectroscopic transition from the H-rich phase to the H-deficient Type
Ib phase, confirming the classification of Type IIb.  Foreground
Galactic reddening, from
\citet{schleg98}\footnote{http://nedwww.ipac.caltech.edu}, is
$E(B-V)=0.01$ or $A_{V}=0.03$ assuming a \citet{ccm89} Galactic
reddening law.  The heliocentric recessional velocity of the host
galaxy is given by \citet{2004AJ....128...16K} as 939\kms.
HyperLEDA\footnote{http://leda.univ-lyon1.fr/} quotes the recessional
velocity, corrected for infall towards the Virgo cluster, to be
$v_{\rm Vir}=758$\kms; for $H_{0}=75$\kms this corresponds to a
distance of 10Mpc.\\ Spectropolarimetric observations of the Type IIb
SN 2001ig were conducted at three epochs: 2001 Dec 16.1, 2002 Jan 3.1
and 2002 Aug 16.3.  These observations were made with the Very Large
Telescope (VLT) at the European Southern Observatory (ESO), Paranal,
Chile; with the Focal Reducer/Low Resolution Spectrograph 1 (FORS1)
instrument \citep{1998Msngr..94....1A}.  A log of these observations
is presented as Table \ref{obstab}.  Throughout this study all times
are a given as UT.  At each epoch FORS1 was mounted on Melipal (Unit
3) VLT at the Cassegrain Focus.  FORS1, in the spectropolarimetry
mode, functions as a dual beam spectropolarimeter.  The FORS1
instrument was used with the standard resolution collimator, providing
a plate scale of $\mathrm{0.2\arcsec\ px^{-1}}$.  The standard
``striped'' PMOS slit mask was used, with slit width 1\arcsec\ and
length 22\arcsec, and for each set of observations the slits were kept
at $\mathrm{PA=0\degr}$.  FORS1 uses a sequence of a ``super
achromatic'' retarder plate and a Wollaston prism (ESO No. 34) to
separate the different polarization components.  Observations of SN
2001ig were conducted with the retarder plate at 4 angles: 0\fdg0,
45\fdg0, 22\fdg5 and 67\fdg5.  At each retarder plate position the
Wollaston prism produced two beams that were dispersed using the grism
G300V, providing a dispersion of 2.6\AA$\mathrm{\ px^{-1}}$ and
spectral resolution, measured from arc lamp calibration exposures, of
12.3\AA.  The observations covered the wavelength range 4450-8600\AA,
when the GG435 order separation filter was used, and 3600-8600\AA,
when no filter was used.  FORS1 uses a $2048\times2048$ Tektronix CCD
detector, and the observations were conducted with gain
$\mathrm{0.71e^{-}ADU^{-1}}$ and readout noise of 5.6$\mathrm{e^{-}}$.
The data were reduced in the standard manner using IRAF\footnote{IRAF
 is distributed by the National Optical Astronomy Observatories,
 which are operated by the Association of Universities for Research
 in Astronomy, Inc., under cooperative agreement with the National
 Science Foundation - http://iraf.noao.edu/} and a series of our own
routines.  The data were corrected for bias and overscan.  A master
unpolarized normalised flat was constructed from flat observations
conducted with the retarder plate at each of the four angles, at each
epoch.  The flat was applied to object frames, and object spectra, for
the ordinary and extraordinary rays at each of the four retarder plate
angles, were optimally extracted.  These spectra were wavelength
calibrated through comparison with observations of HeHgArCd arc lamps.
The Stokes parameters $Q$ and $U$ and total polarization $p$ and
polarization angle $\theta$ were calculated in the standard manner
\citep[see][]{forsman}, with the data re-binned to 15\AA\ to improve
the signal-to-noise (S/N) for the individual Stokes parameters.  Here
we have used a flux binning technique, with bins recalculated prior to
the calculation of the Stokes parameters.  This yields identical
results to the weighted Stokes parameter binning of
\citet{1997ApJ...476L..27W}.  The bin size was determined by the
spectral resolution of the observations, and was required to be much
less than the velocity widths of the SN features
\citep{2003ApJ...591.1110W}.  A correction was applied for the
wavelength dependent chromatic zero angle offset.  The total flux
spectra of SN 2001ig were flux calibrated using observations of flux
standard stars with the retarder plate set to 0\fdg0.  The data were
corrected for the heliocentric recessional velocity of the host
galaxy.
\section{Observational Results}
\label{results}

\subsection{Flux Spectra} 
\label{spectrores}
\begin{figure}
\hfil
\rotatebox{-90}{
\includegraphics[width=15cm]{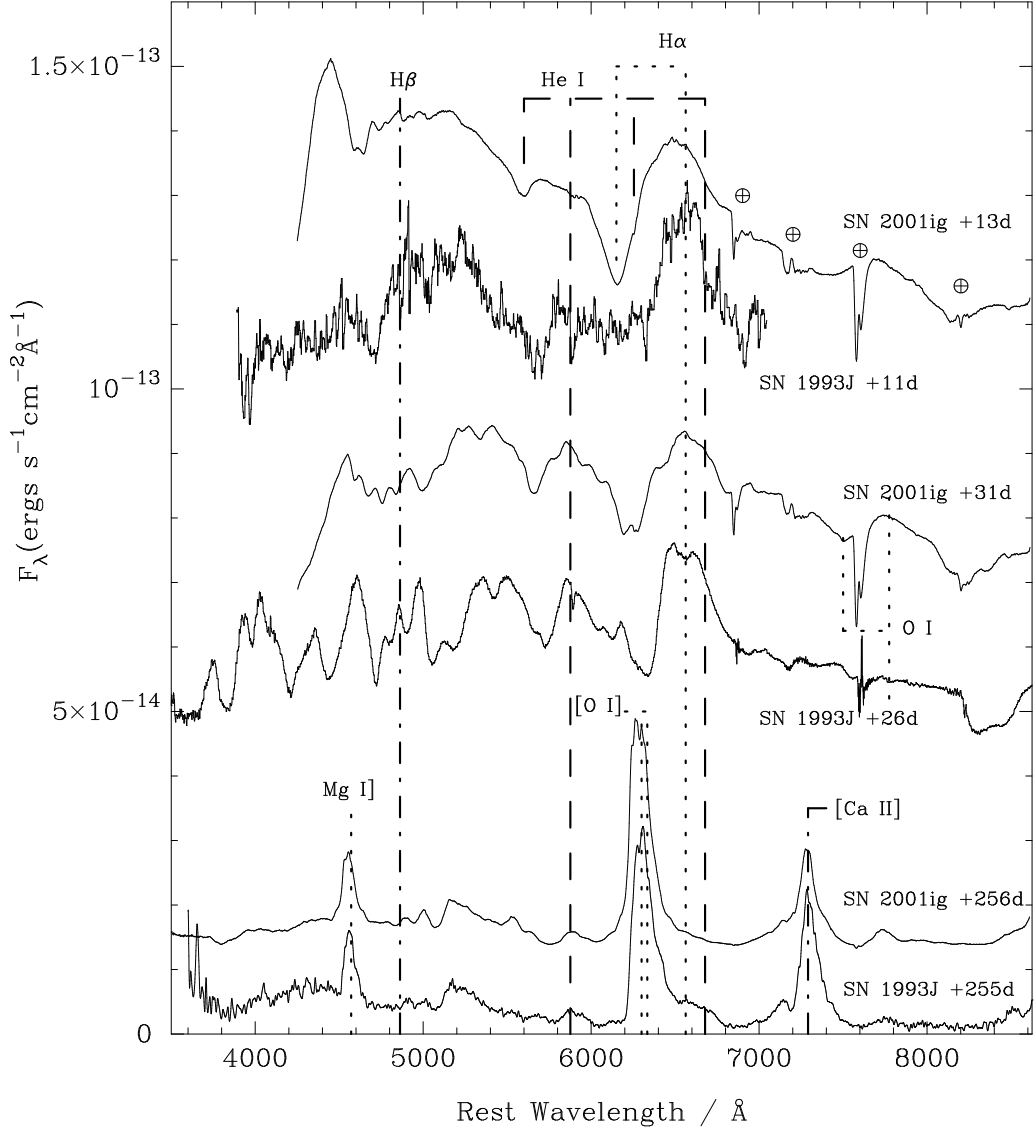}}
\hfil
\caption{Flux spectra of SN 2001ig at 2001 Dec 16, 2002 Jan 3 and 2002 Aug 16 (or 13, 31 and 256 days post-explosion respectively).  Line identifications follow \citet{2000AJ....120.1487M}.  Telluric lines are indicated by the $\mathrm{\oplus}$ symbol.  Also shown are spectra of SN 1993J, from the SUSPECT archive, similarly scaled for similar post-explosion epochs, as given on the figure, using the explosion date of SN 1993J of \citet{1997PASP..109..489T}.  The vertical lines show the rest wavelengths and, where appropriate, the associated absorption profile.  The wavelengths of these spectra have been corrected for the recessional velocities of the host galaxies of these two SNe.}
\label{obsres:flux}
\end{figure}

Flux spectra of SN 2001ig, at the three epochs, are presented as
Fig. \ref{obsres:flux}. Velocities at the absorption minima for
different lines, in the spectra of 2001 Dec 16 and 2002 Jan 03, are
given in Table \ref{linestab}.\\ The spectrum of 2001 Dec 16, at 13
days post-explosion, is dominated by the strong \halpha\ P Cygni
profile, with an absorption minimum at $-19\,000$\ \kms.  The peculiar
profile of the line, with sharp edges separating the emission and
absorption components, and a redward shoulder to the emission
component suggests blending with He I 6678\AA.  This interpretation is
further validated by the presence of an absorption ``notch'' at
6255\AA, corresponding to He I 6678\AA\ also blue shifted by
$-19\,000$\ \kms.  There may also be a contribution to the \halpha\
profile due to blended Si II 6355\AA, which is seen in the spectrum of
the next epoch's observation, but this is not resolved at the first
epoch.  An absorption at 5603\AA\ is identified as He I 5876\AA\ (-14
000\ \kms), however the associated emission feature is not symmetrical
about the rest wavelength, the peak is blue ward of the rest
wavelength.  It may be that the emission feature is truncated by some
other feature that lies redward.  An absorption feature at the red end
of the spectrum is identified as the absorption associated with the Ca
II IR triplet ($\sim -12\,700$\ \kms) - with the emission feature
redward of the wavelength limit of the observation. This absorption
is, however, blended with a broad telluric feature, as indicated on
Fig. \ref{obsres:flux}, increasing the uncertainty on the measured
velocity at the true absorption minimum.  There is a single absorption
at the rest wavelength of Na ID at the recessional velocity of the
host galaxy.  The measured equivalent width of this feature was
determined to be $W_{\lambda}(\mathrm{Na ID})=0.7\pm0.1$\AA.  Using
the relations of \citet{2003fthp.conf..200T}, this equivalent width
corresponds to two possible reddenings of $E(B-V)=0.10$ or $0.32$,
which are due to internal reddening in the host galaxy.\\ The spectrum
of 2002 Jan 3, $31$ days post-explosion, shows the SN at a
transitional state from a Type II to Type Ib SN, with the He lines
growing in strength compared to the hydrogen lines.  The contamination
of \halpha\ by lines from other species is more evident.  The sharp
edge between the absorption and emission components of this profile
results from the superposition of Si II 6355\ang, \halpha\ and He I
6678\ang.  As in the previous epoch, the He I emission component
produces a redward shoulder on the \halpha\ emission component, and
leads to an absorption ``notch'' at 6415\ang (or $-12\;000$\ \kms)
which suppresses the blue side of the \halpha\ emission. The
absorption minimum blue ward of the \halpha\ peak is characterised by
two minima at 6192\ang and 6265\ang, which corresponds to the Si II
($-8\;000$\ \kms) and \halpha\ ($-13\;600$\ \kms), respectively.  He I
5876\ang is observed as a much stronger feature, with obvious emission
and absorption components; the absorption minimum corresponds to
$-12\;000$\ \kms.  The steep transition blueward from the He I emission
to the absorption shows a sharp notch due to Na ID.  O I 7774\AA\ is
present as a P Cygni profile with an absorption minimum at 7500\AA,
corresponding to $-10\,600$\kms.\\ The final spectrum, of 2002 Aug 16,
$256$ days post-explosion, shows the complete transformation of the SN
spectrum to the nebular phase, dominated by strong forbidden line
emission features, typical of other CCSNe at the nebular phase
(e.g. SN 1993J; \citealt{2000AJ....120.1499M}).  Most prominent at
this epoch is the O I feature at $\mathrm{\lambda\lambda 6300,
6364}$\ang, which shows a number of peaks that are indicative of
asymmetries in the expanding ejecta.  A crude Gaussian fit to the O I
emission feature gives a full width at half maximum of $6\;700$\kms.
Also present in the spectrum are features due to Mg I] 4571\AA, [Ca
II] 7291,7324\AA, and emission features due to Fe II 5169\AA, HeI
5876\AA\ and O I 7774\AA.\\

\subsection{Spectropolarimetric Properties of SN 2001ig}
\label{specpolres}
\subsubsection{General Spectropolarimetric Properties}
\label{genspecpol}
The spectropolarimetric observations of SN 2001ig, at the three
epochs, are presented as Figs. \ref{obsres:main:epoch1},
\ref{obsres:main:epoch2} and \ref{obsres:main:epoch3}.  The observed
polarization properties are a product of both the intrinsic
polarization of the SN itself and polarization due to the intervening
Interstellar Medium (ISM).  The calculation of the latter, the
Interstellar Polarization (ISP), is an important consideration and the
determination of its quantity is discussed at length in the following
sections along with analysis of the intrinsic polarization of the SN.
Importantly, the change in the polarization properties between 2001
Dec 16 and 2002 Jan 3 is directly correlated with the evolution in the
spectroscopic features in the SN spectrum.  The polarization features
associated with He I show particularly obvious evolution, with
increasing polarization as the He I flux line strength increases at
the later epoch.  In addition, we note that the observed polarization
P.A. in the first epoch varies widely across $\sim45\degr$ in the
first epoch, whereas in the second epoch the P.A. falls within a
narrower range of $\sim 20\degr$.  The polarization features in the
second epoch are directly correlated with spectral lines, and for
instances with P Cygni profiles we see the classic ``inverted P
Cygni'' profiles in the polarization spectra
\citep{1988MNRAS.231..695C}.  These profiles are observed for \halpha
, He I 6678\AA\ and O I 7774\AA.\\ The observations of the last epoch
(2002 Aug 16 or 256 days post-explosion) show that, with the SN having
entered the nebular phase, the spectrum is almost completely
unpolarized, without significant correlation between polarization
features and spectral features.  At such late epochs it is presumed
that geometric dilution of the scattering particles in the SN ejecta
leads to a drop in the scattering optical depth such that, even in the
presence of asymmetries, the density of scatterers is not sufficiently
high to produce the polarization signature.  The amount of scatter in
the polarization properties of neighbouring wavelength bins is
directly correlated with the S/N and in instances where the S/N is
high, as for the $\mathrm{\lambda4571}$ Mg I],
$\mathrm{\lambda\lambda6300,6334\AA}$\ [O I] lines and
$\mathrm{\lambda\lambda7291,7324}$ [Ca II] lines, the data has small
uncertainties and have polarizations tending towards the same values
(see \S\ref{isp}).

\subsubsection{Data on the Q-U plane}
\label{qu}
The $Q-U$ plane is a powerful tool for studying the effects of the ISP
and simultaneously examining the behaviours of the polarization and
the polarization angle with wavelength.  This tool has been used to
great effect by a number of studies
\citep{1988MNRAS.231..695C,2003ApJ...591.1110W,2003ApJ...592..457W}.
\citet{2003ApJ...592..457W} showed how the distribution of the data on
the Q-U plane can be parameterised by a ``dominant axis,'' with
functional form:
\begin{equation} 
U=\alpha+\beta Q
\label{eqtn:dom1}
\end{equation}
This axis can be used to study and quantify symmetries in the
polarization data, which are directly linked to physical axial
symmetries in the SN.  In addition, \citet{2006ApJ...653..490W} showed
that $Q$ and $U$ Stokes parameters across spectral lines have their
own ``dominant'' axes, leading to information about the relative
distribution of particular species.  \citet{1988MNRAS.231..695C}
showed that, for spectropolarimetry of $\mathrm{H\alpha}$ of SN 1987A,
the progression through the points on the $Q-U$ plane, in wavelength
order, yielded a loop structure, as opposed to a simple straight line.
This was interpreted as the symmetry axis, of the aspherical SN,
having different orientations as a function of wavelength \citep[see
also][]{2003ApJ...593..788K}.  The distribution of the
spectropolarimetry data of SN 2001ig on the $Q-U$ plane are shown, for
each of the three observational epochs, as
Figs. \ref{obsres:qu:epoch1}, \ref{obsres:qu:epoch2} and
\ref{obsres:qu:epoch3}.\\ Since all the observations have been
conducted with the slits in the same orientation ($PA=0\degr$), all
the polarization data presented here has Stokes parameters in the
reference frame of the sky.  At the first epoch, the concentration of
the data on the $Q-U$ plane is obvious, although the location of the
dominant axis is not immediately clear.  The dominant axis, at each
epoch, was calculated using a $\chi^{2}_{\nu}$ minimization technique,
that excluded the low S/N red end of the spectrum corresponding to the
Ca II IR triplet absorption.  The low intrinsic polarization data at
the blue end of the spectrum ($\mathrm{<6000\AA}$) lies on one side of
the dominant axis, while the data associated with the \halpha\ feature
lie on the opposite side of the dominant axis (as shown on
Fig. \ref{obsres:qu:epoch1}).  At the second epoch the dominant axis
is clearly defined and has rotated through $\sim40\degr$ compared to
the first epoch.  The data is more tightly clustered on the $Q-U$
plane, which is partially due to an improvement in the S/N at the
second epoch as the SN had just reached maximum by day 31.  The
difference in the locus of the data, on the $Q-U$ plane, at the two
different epochs results in two different values of $\beta$ (even
after ISP subtraction, as discussed in Section \ref{isp}).  The data
at day 256 show no discernible structure and are dominated by the
statistical uncertainty due to the faintness of the SN, compared to
earlier epochs, and, hence, low S/N.  The use of a dominant axis to
describe the data at the third epoch was, therefore, inappropriate.\\
The dominant axis defines a useful ``zero point'' for rotating the
Stokes Parameters to a more meaningful reference frame than that
defined by using $PA=0\degr$ on the sky.  The $Q$ and $U$ parameters
can be rotated using the standard two-dimensional rotation
transformation, given by \citep{2003ApJ...591.1110W} as:
\begin{equation}
P_{d}=(Q-Q_{ISP})\cos 2\theta_{d}+(U-U_{ISP})\sin 2\theta_{d}
\label{eqtn:dom2}
\end{equation}
\begin{equation}
P_{o}=-(Q-Q_{ISP})\sin 2\theta_{d}+(U-U_{ISP})\cos 2\theta_{d}
\label{eqtn:dom3}
\end{equation}
where $\theta_{d}$ is the polarization angle of the dominant axis,
determined from Eqn. \ref{eqtn:dom1}, and the $P_{o}$ is the polarization in the orthogonal direction on the $Q-U$ plane but corresponds to a physical rotation of $45\degr$.  To ensure rotation about an
origin of zero intrinsic polarization of the SN, the rotated Stokes
parameters are calculated using the Stokes parameters corrected for
the ISP (see \S\ref{isp}).  The rotated Stokes parameters for the data
of 2001 Dec 16 and 2002 Jan 03 are shown as
Figs. \ref{obsres:dom:epoch1} and \ref{obsres:dom:epoch2},
respectively.
\subsubsection{Determination of the Interstellar Polarization}
\label{isp}
\begin{figure}
\hfil
\rotatebox{-90}{
\includegraphics[width=15cm]{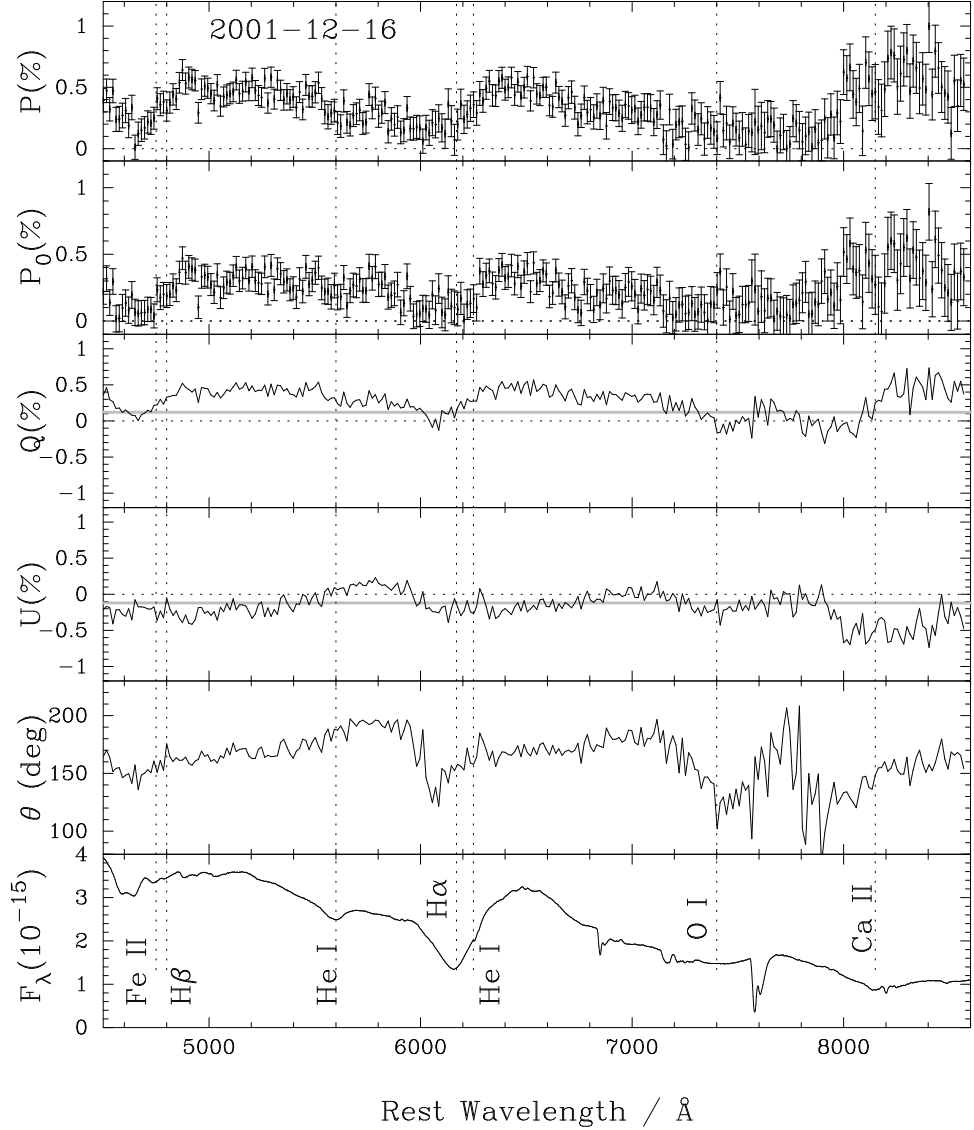}}
\hfil
\caption{Spectropolarimetry of SN 2001ig, acquired on 2001 Dec 16 (13 days post-explosion), showing the total observed polarization $p$, the total polarization after correction for the ISP $p_{0}$, the Stokes $Q$ and $U$ parameters, the polarization angle $\theta$ and the total flux spectrum ($\mathrm{ergs\,s^{-1}\,cm^{-2}\,\AA^{-1}}$).  The Stokes parameters have been re-binned to 15$\mathrm{\AA}$.  The thick grey lines indicate the position of the ISP Stokes $Q$ and $U$ parameters, effectively the zero-point for the Stokes parameters intrinsic the SN.  The zeropoint for observed $Q$ and $U$ is indicated by the horizontal dotted line.  All data is presented with the wavelength scale corrected for the heliocentric recessional velocity of the host galaxy.}
\label{obsres:main:epoch1}
\end{figure}

The correct study of the intrinsic polarization properties of SN
2001ig requires careful accounting for and subtraction of polarization
due to the intervening ISM.  \citet{1975ApJ...196..261S} present the
relationship between reddening and the ISP, where $p_{ISP}(\%) <
9E(B-V)$ for Galactic type reddening and dust polarization laws.  This
relationship can be used to place important maximum limits of the
total $B/V$ polarization using reddenings determined from
non-polarimetric techniques.  We assume that dust in the host galaxy
NGC 7424 follows similar reddening and polarization laws as the
Galaxy, although studies by \citet{2002AJ....124.2506L} and
\citet{2004AJ....127.3382C} suggest that this may not be the case.
The maximum polarization due to the Galactic ISM is $0.09\%$, given
$E(B-V)=0.01$.  In addition, the reddening calculated from the
equivalent width of the Na ID lines, in \S\ref{spectrores}, yields two
limits on the polarization, due to the ISM of the host galaxy, as
0.9\% and 2.9\%.  Under the assumption that the Galactic polarization
law holds in NGC 7424, these are strict upper limits on the ISP
(including the relevant uncertainties in the measurements of the
equivalent width and in the calculation of the reddening).  These
limits describe a circle on the $Q-U$ plane, and comparison of
Figs. \ref{obsres:qu:epoch1} to \ref{obsres:qu:epoch3} shows that the
observed data have polarizations significantly lower than these
limits.\\ \citet{1993ApJ...414L..21T} and \citet{1997PASP..109..489T}
assume that the P Cygni emission components of lines such as \halpha\
are inherently depolarized, such that any residual polarization across
the emission features is due solely to the ISP.  At the first epoch,
Figs. \ref{obsres:main:epoch1} and \ref{obsres:qu:epoch1}, the
\halpha\ emission component has one of the highest polarization levels
across the observed wavelengths.  The \halpha\ profile, as discussed
in Section \ref{spectrores}, is a product of both \halpha\ emission
and the blueward absorption associated with the He I 6678\AA. In this
case, therefore, the polarization observed across \halpha\ is
obviously not the ideal case of a completely depolarized emission
component of one line and, hence, precludes the use of this assumption
for determining the ISP.\\ As we have observations at more than one
epoch, it is possible to calculate the ISP as a non-varying quantity,
with a single polarization angle, by comparing the behaviour of SN
2001ig on the $Q-U$ plane at different epochs.  There could also be a
local,variable, polarized source within the seeing disk of these
observations; there is no known technique to correct for such
phenomena.  As noted in \S\ref{genspecpol}, the data show a large
rotation and an increase in polarization between the first and second
epochs, corresponding to the change from an H-rich spectrum to a
transitional He-rich spectrum.  The ISP is, by assumption, constant
between the two epochs, requiring our choice of ISP to be applicable
to both cases.  Furthermore, following \citet{2001ApJ...556..302H} and
\citet{2003ApJ...591.1110W}, we expect the ISP to be located at one
end of the dominant axis of the locus of the data on the $Q-U$ plane
and in the proximity of the data at bluer wavelengths, due to high
degree of blending and, hence, depolarization of Fe II lines at bluer
wavelengths.  For the data of the first epoch
(Fig. \ref{obsres:qu:epoch1}) we note that the $Q$ and $U$ of the
lowest total polarization features limit the ISP to $Q_{ISP}<0.2$,
$U_{ISP}>-0.2$.  For the second epoch, shown as
Fig. \ref{obsres:qu:epoch2}, the dominant axis is easily identified,
and the ISP must be located at the low polarization end of the
distribution of data along the dominant axis (since an ISP located at
the opposite high-polarization end of the data distribution would be
inconsistent with the observed data of the first epoch).  The data of
the second epoch limits the ISP Stokes parameters to $Q_{ISP}<0.3\%$
and $U_{ISP}<0.1\%$.\\ The observation at the third epoch shows the SN
in the nebular phase, when the density of scattering particles is
extremely low.  At this epoch it is expected that the density is
insufficient to produce a net intrinsic polarization, rather that the
observed polarization at this epoch is just due to the combined ISP of
the foreground and host galaxy ISP.  The average values of $Q$ and $U$
data would therefore be equivalent to $Q_{ISP}$ and $U_{ISP}$.  In
order to use this data to determine an ISP, the data were rebinned to
50\AA\ to improve the S/N.  The polarization signature at the red end
of the data suffers from low S/N, due to the response function of
FORS1 at those wavelengths.  This leads to a polarization signature
with larger uncertainties.  This section of the spectrum, with
$\mathrm{\lambda > 8000\AA}$, was removed from the calculation.  The
rebinning of the spectrum, and removal of high-uncertainty wavelength
range, significantly reduced the scatter in the data and the centroid
of the data on the $Q-U$ plane was measured, using separate weighted
averages for both $Q$ and $U$, to be $\overline{Q}=0.12\pm0.07$ and
$\overline{U}=-0.12\pm0.08$.  The quoted uncertainties on these values
are the standard deviations, which provide a better estimate of the
scatter than the standard error on the mean.  These values are
consistent with the limits placed on them using the observations from
the first and second epochs, and are the same as the measured Stokes
parameters in the regions of the spectrum of the third epoch with the
highest S/N (see \S\ref{genspecpol}).\\ 
The \citet{1975ApJ...196..261S} form of the ISP has only a small
degree of $\lambda$-dependence, across the wavelength range being
studied here.  The small value of the determined Stokes parameters of
the ISP are sufficiently low that any effects due to the
$\lambda$-dependence of the ISP would be much less than the quoted
uncertainties.  We adopt, therefore, values of the ISP Stokes
parameters of $Q_{ISP}=0.12\pm0.07$ and $U_{ISP}=-0.12\pm0.08$,
corresponding to $p_{ISP}=0.17\pm0.08$ and $\theta_{ISP}=157$\fdg$5$
(which is in the direction tangential to the host spiral arm in NGC
7424, as shown on Fig. \ref{fig01igpos};
\citealt{1987MNRAS.224..299S}).  These values limit the reddening,
assuming a Galactic polarization law, to $E(B-V)>0.01$ as a lower
limit, requiring a reddening component in addition to the Galactic
foreground reddening.\\
Correction of the ISP requires the vectorial subtraction of $Q$ and
$U$, effectively an offset of the zero-point of the intrinsic $Q$ and
$U$ parameters from the observed Stokes parameters.  The principal
ramification of this property is that the corrected total polarization
$p_{0}\neq p_{obs}-p_{ISP}$, and polarized features can become
depolarized and vice versa.  The polarization spectra, after
correction for the ISP, are also shown for the three observational
epochs as Figs. \ref{obsres:main:epoch1}, \ref{obsres:main:epoch2} and
\ref{obsres:main:epoch3}.\\ \citet{2001ApJ...556..302H} suggest that
the region between 4800-5600\AA\ should be depolarizing, since in that
wavelength range the line blanketing opacity of Fe II lines dominates
over electron scattering, leading to net depolarization.  We note that
in the case of SN 2001ig, however, the observed polarization over much
of that wavelength range is similar in magnitude to that observed for
\halpha\, for which the presence of net intrinsic polarization has
already been discussed.  The narrower wavelength region of
4600-4800\AA\ shows lower levels of polarization, at the first epoch,
than 4800-5600\AA.  Only in the narrower region 4600-4800\AA\ is there
almost complete depolarization evident, with the observed polarization
consistent with the ISP alone (see Fig. \ref{obsres:qu:epoch1}).  The
Fe II lines are further discussed in \S\ref{feiisec}.\\
\subsubsection{\halpha, Si II and He I 6678\AA}
\label{analysishalpha}
\begin{figure}
\hfil
\rotatebox{-90}{
\includegraphics[width=15cm]{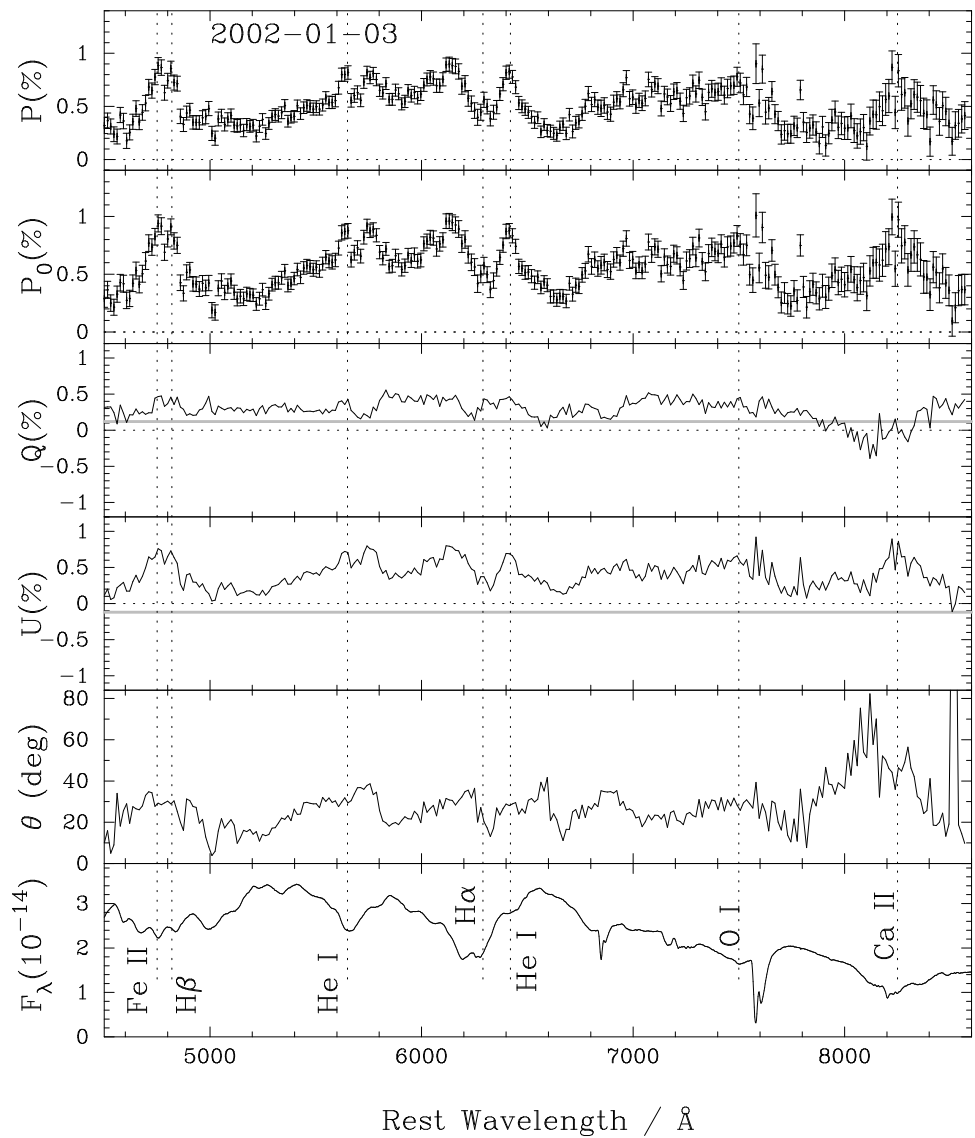}}
\hfil
\caption{Same as Fig. \ref{obsres:main:epoch1}, but for spectropolarimetry of SN 2001ig acquired on 2002 Jan 3 (31 days post-explosion).  The data is uncorrected for the ISP.  Note the small dispersion of the polarization angle $\theta$, compared to the earlier epoch of Fig. \ref{obsres:main:epoch1}.}
\label{obsres:main:epoch2}
\end{figure}

\begin{figure}
\hfil
\rotatebox{-90}{
\includegraphics[width=15cm]{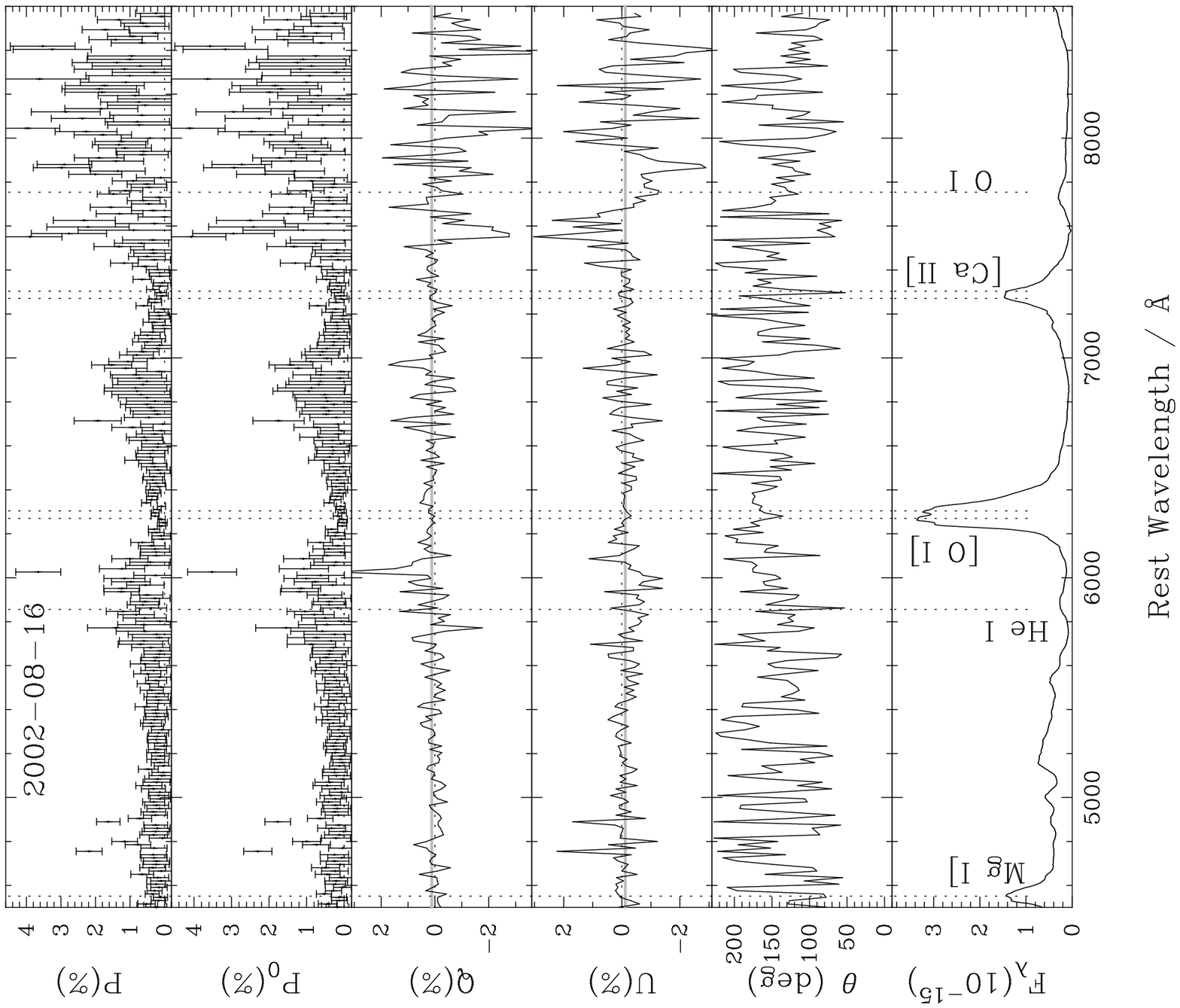}}
\hfil
\caption{Same as Fig. \ref{obsres:main:epoch1}, but for spectropolarimetry of SN 2001ig acquired on 2002 Aug 16 (256 days post-explosion).}
\label{obsres:main:epoch3}
\end{figure}

\begin{figure}
\hfil
\rotatebox{-90}{
\includegraphics[width=10cm]{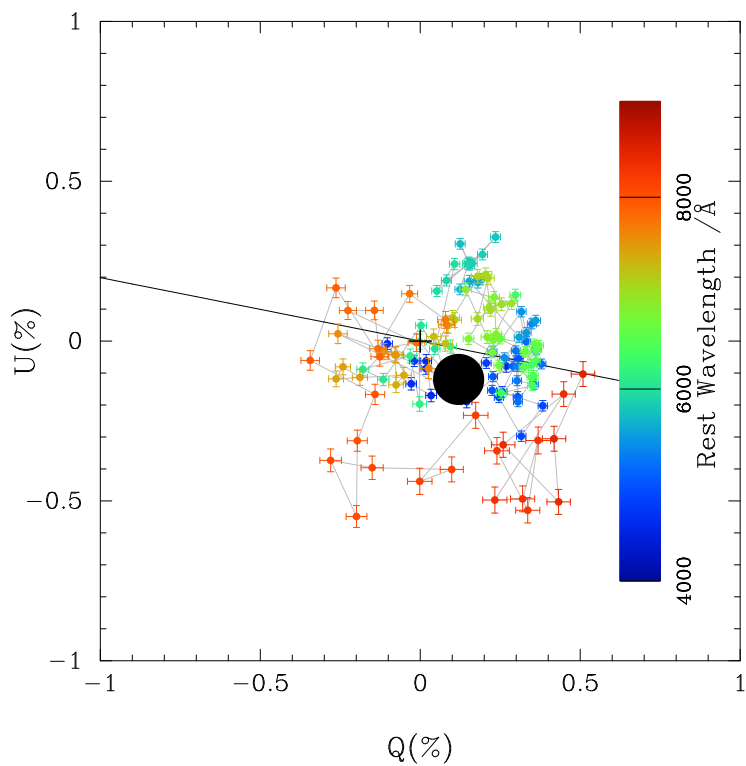}}
\hfil
\caption{Stokes $Q$ and $U$ parameters, as a function of wavelength from observations acquired on 2001 Dec 16 (13 days post-explosion).  The Stokes parameters have been re-binned to 50$\mathrm{\AA}$.  The data have been corrected for the ISP (shown as the solid black circle; see text).  The dominant axis (see text) is indicated by the straight line.  The wavelength of each point is indicated by the colour, following the scheme of the colour bar on the right hand side.  The location of the origin of the Stokes plane ($Q=0$ and $U=0$) is indicated by cross.}
\label{obsres:qu:epoch1}
\end{figure}

\begin{figure}
\hfil
\rotatebox{-90}{
\includegraphics[width=10cm]{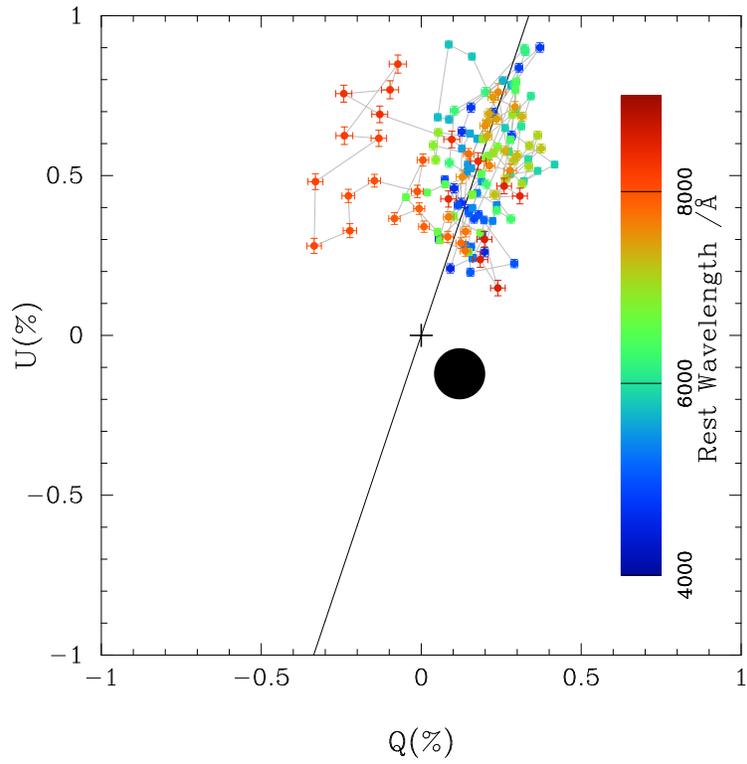}}
\hfil
\caption{Same as Fig. \ref{obsres:qu:epoch1}, but for
spectropolarimetry of SN 2001ig acquired on 2002 Jan 3 (31 days
post-explosion).  Note that the dominant axis has rotated {\it physically}
$\sim40\degr$ compared to that of the earlier epoch of
Fig. \ref{obsres:qu:epoch1}.}
\label{obsres:qu:epoch2}
\end{figure}

\begin{figure}
\hfil
\rotatebox{-90}{
\includegraphics[width=10cm]{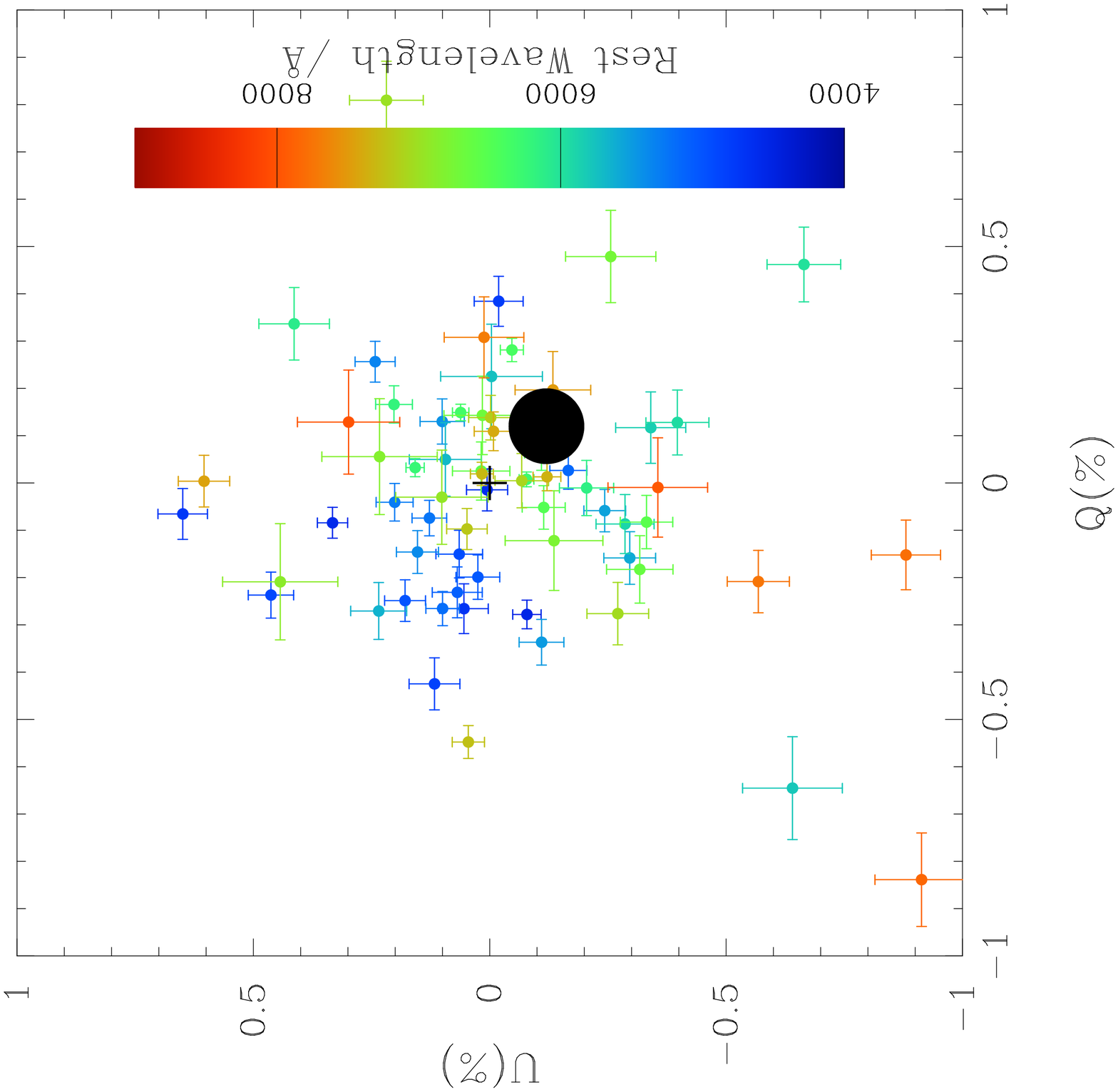}}
\hfil
\caption{Same as Fig. \ref{obsres:qu:epoch1}, but for spectropolarimetry of SN 2001ig acquired on 2002 Aug 16 (256 days post-explosion).}
\label{obsres:qu:epoch3}
\end{figure}

\begin{figure}
\hfil
\rotatebox{-90}{
\includegraphics[width=10cm]{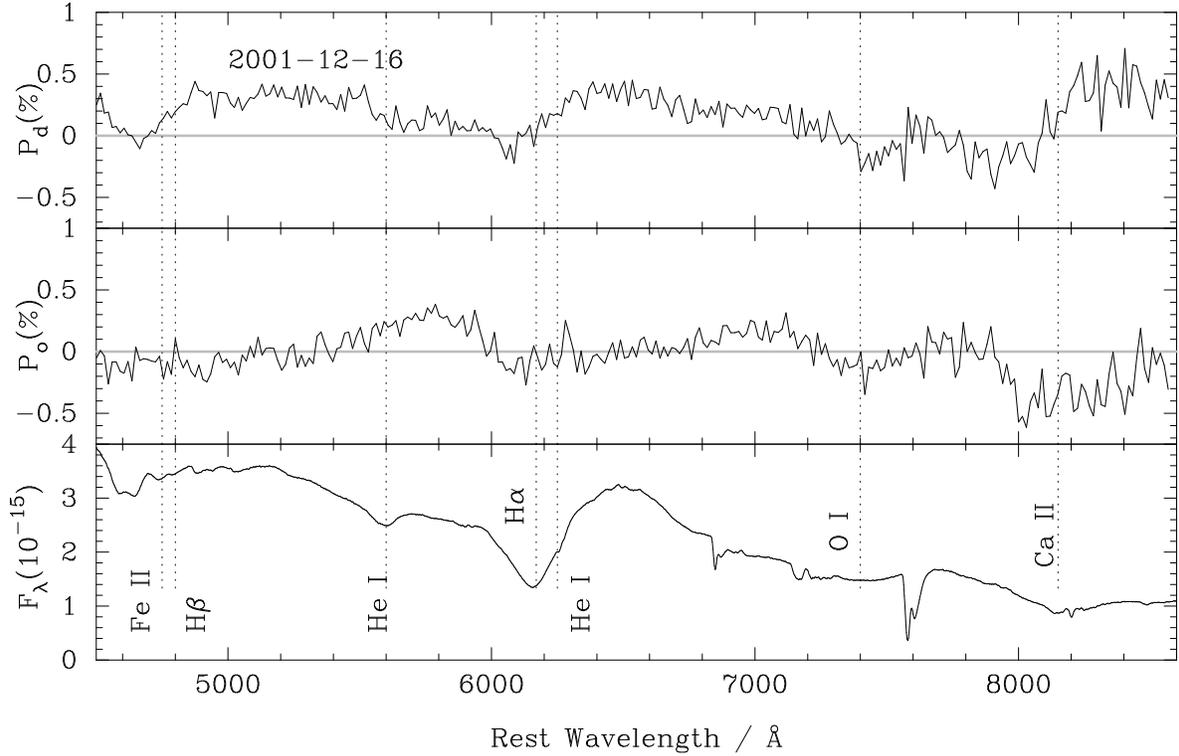}}
\hfil
\caption{The Stokes parameters, corrected for the ISP, rotated to the dominant ($P_{d}$) and orthogonal axes ($P_{o}$) for the data of 2001 Dec 16, compared with the flux spectrum at this epoch.  There is little difference between these rotated Stokes parameters and the $Q$ and $U$ Stokes parameters (Fig. \ref{obsres:main:epoch1}), since the dominant axis is nearly aligned ($\theta_{d}=4.5\degr$) with the $Q$ axis.  Zero polarization is indicated by the heavy grey line.}
\label{obsres:dom:epoch1}
\end{figure}

\begin{figure}
\hfil
\rotatebox{-90}{
\includegraphics[width=10cm]{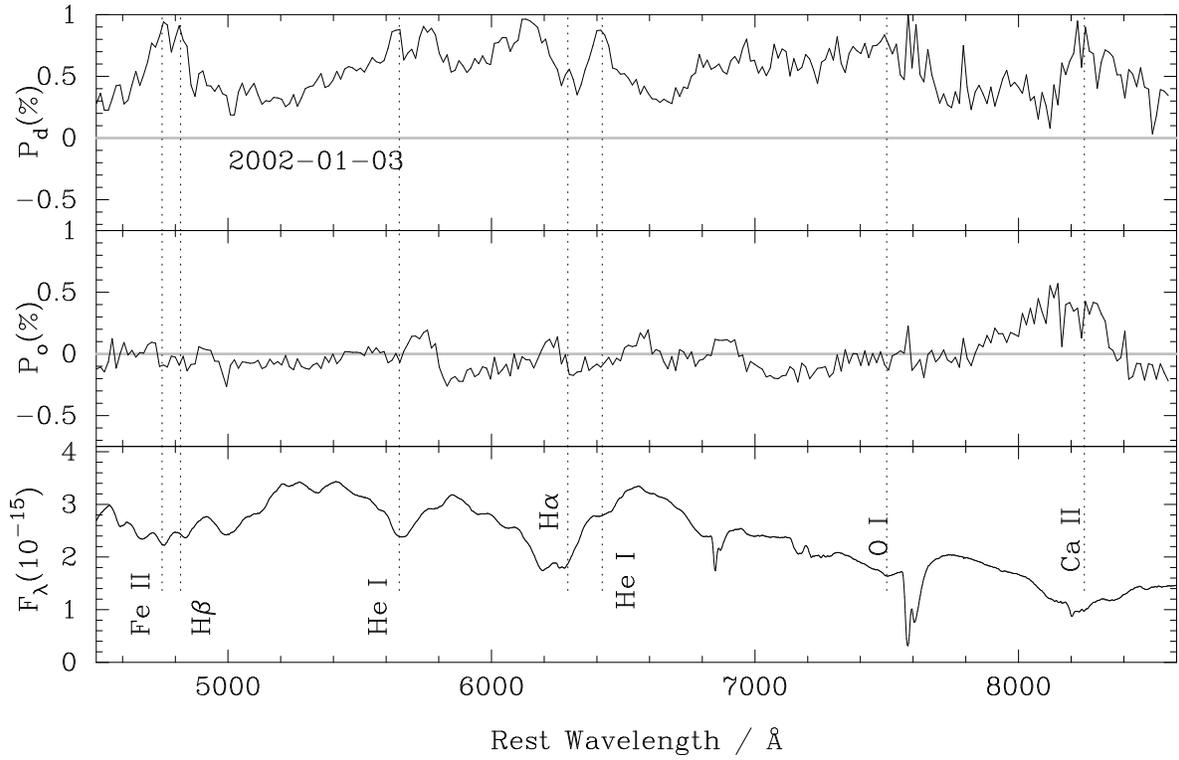}}
\hfil
\caption{The Stokes parameters, corrected for the ISP, rotated to the dominant ($P_{d}$) and orthogonal axes ($P_{o}$) for the data of 2002 Jan 03, compared with the flux spectrum at this epoch. Zero polarization is indicated by the heavy grey line.}
\label{obsres:dom:epoch2}
\end{figure}

\begin{figure}
\figurenum{11a}
\hfil
\rotatebox{-90}{
\includegraphics[width=10cm]{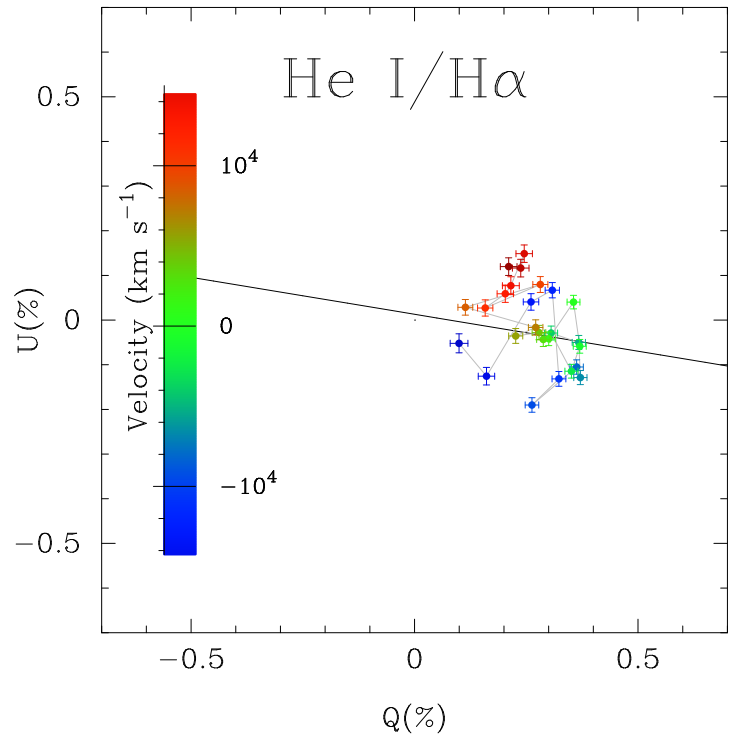}}
\hfil
\caption{Loops on the Q-U plane for \halpha/He I 6678\AA\ blend, the wavelength range covering O I 7774\AA, Fe II lines in the range 4800-5600\AA,\ and Ca II IR absorption from the observations of 2001 Dec 16.  The data have been re-binned to 15\AA.  The heavy dashed line indicates the dominant axis, calculated for the entire data set for this epoch shown as shown on Fig. \ref{obsres:qu:epoch1}.}
\label{disc:quloop:epoch1:a}
\end{figure}

At the first epoch, the \halpha\ polarization signature seems
counter-intuitive, with the maximum polarization of this feature
observed at 6350\AA, near the emission peak in the flux spectrum, and
minimum polarization reached just blueward of the absorption minimum.
Low polarization is generally seen for the entire absorption feature
(see Fig. \ref{obsres:main:epoch1}), with the Stokes parameters lying
on or close to the dominant axis with no obvious loop structure
(Figs. \ref{obsres:dom:epoch1} and \ref{disc:quloop:epoch1:a}).  At day
31, the blue edge of the \halpha/Si II absorption profile
($\sim6100$\AA) is coincident with a peak in polarization of 0.7\%
(Figs. \ref{obsres:main:epoch2} and \ref{obsres:dom:epoch2}).
Significant polarization is observed at both of the flux minima of
\halpha\ and Si II, but it is smaller in magnitude than the
polarization at the blue edge.  Redward of the flux absorption minima
is a partly depolarized region at 6300\AA\ associated with the
\halpha\ emission.  The polarization has a local minimum of 0.2\% at
the location of the He I 6678\AA\ emission component.  Within this
depolarized wavelength range there is, however, a peak in polarization
at $\sim6400$\AA, comparable to that observed for the blue edge of the
\halpha/Si II absorption profile.  This peak coincides with the
absorption minimum of the He I 6678\AA\ P Cygni profile, the prominent
``notch'' at 6400\AA\ in the flux spectrum
(Figs. \ref{obsres:main:epoch2} and \ref{obsres:dom:epoch2}).  This
polarization signature across the He I 6678\AA\ absorption minimum
matches the polarization peak observed for the absorption minimum of
He I 5876\AA, where the peak polarizations are equal and the Stokes
Parameters are observed to rotate through 10$\degr$.\\ The behaviour
of the blended \halpha/He I/Si II feature on the Q-U plane, presented
in Fig. \ref{disc:quloop:a}, shows that data at these wavelengths are
aligned with the dominant axis, but that the H I absorption feature
forms a loop deviating from a straight line by $\pm0.2\%$, in the
direction orthogonal to the dominant axis.  The rotated Stokes
parameters at this epoch (Fig. \ref{obsres:dom:epoch2}) actually
resolve the separate structures of the blend.  A peak in the
orthogonal rotated Stokes parameter is associated with the absorption
minimum of \halpha\ in the flux spectrum, but there is no
corresponding feature in $P_{d}$.  The absorption feature associated
with He I 6678\AA\, giving rise to the ``notch'' in the blended
profile in the flux spectrum, is, conversely, only observed in $P_{d}$
and not in $P_{o}$.  The loop structure observed on the $Q-U$ plane
for this blend is, therefore, due to the superposition of two lines
arising from different species with different geometries within the
ejecta.  By day 256, \halpha\ nebular emission is not significantly
detected in the flux spectrum and there is no significant polarization
at the expected wavelength for this line.
\subsubsection{O I 7774\AA}
\begin{figure}
\figurenum{11b}
\hfil
\rotatebox{-90}{
\includegraphics[width=10cm]{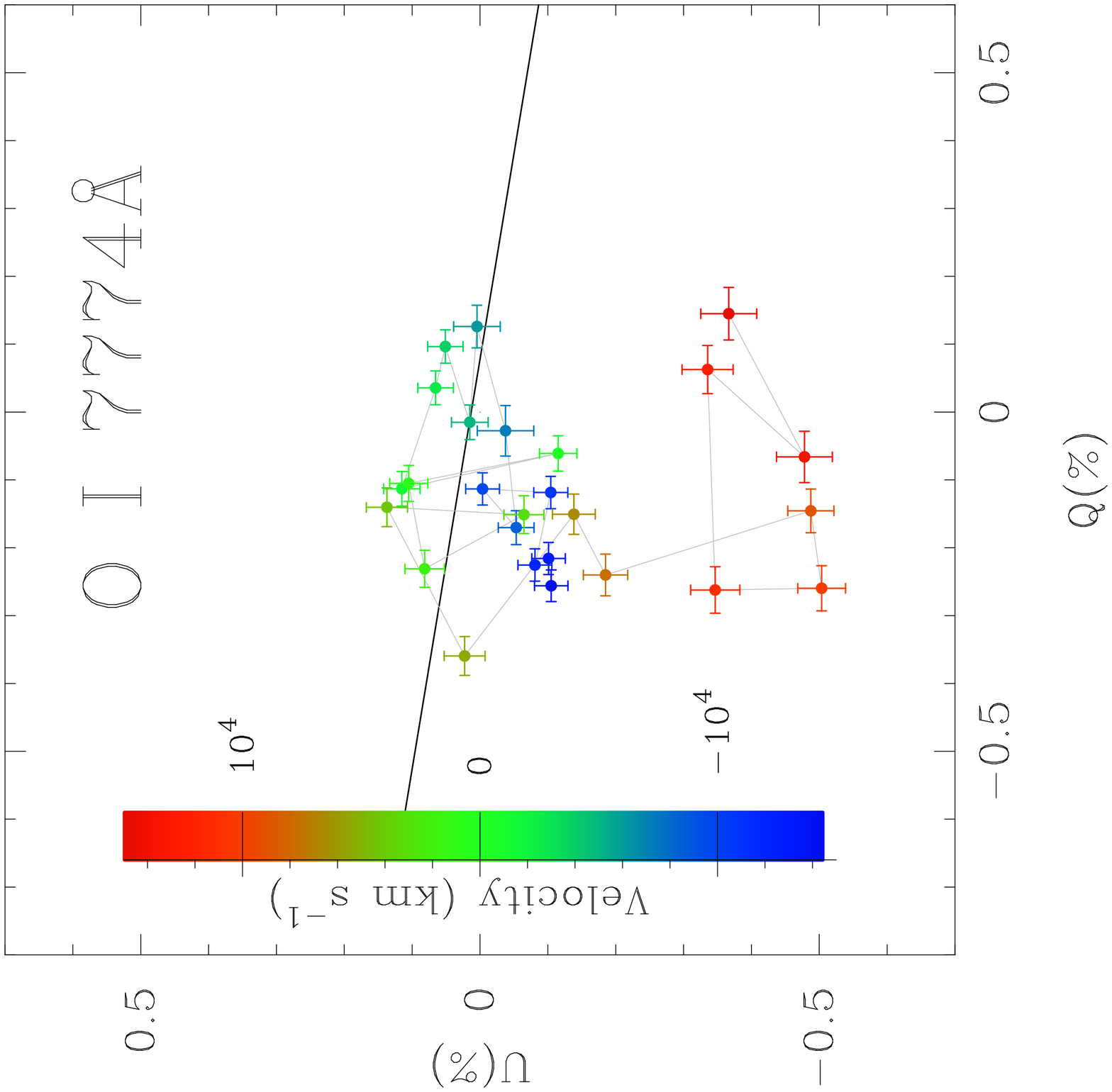}}
\hfil
\caption{}
\label{disc:quloop:epoch1:b}
\end{figure}

The O I 7774\AA\ feature is observed in the flux spectra at all
epochs, and in the case of the last epoch it is only seen in emission.
At the first epoch there is no excess polarization, over the
wavelength range of the O I 7774\AA\ line (which is observed at the
second epoch) as shown in Fig. \ref{disc:quloop:epoch1:b}.  At 2002
Jan 3, the standard polarization profile expected for P Cygni
spectroscopic profiles is observed, with peak polarization (0.7\%)
associated with the absorption component at $\sim7500$\AA\ and
decreasing levels of polarization to a minimum of 0.3\% at the peak of
the emission component (Fig. \ref{obsres:main:epoch2}).  These maxima
and minima are similar to the observed polarization levels of \halpha\
and He I 6678\AA.  The presence of a telluric absorption band at the
wavelengths of this feature complicates both the spectroscopic and
spectropolarimetric interpretation of this line.  There is sufficient
resolution, even with the data rebinned, to safely identify those bins
with high uncertainties on the Stokes parameters and exclude them from
the analysis.  On the $Q-U$ plane, on Fig. \ref{disc:quloop:b}, the O I
data points are tightly clustered about the dominant axis, with a
maximum deviation in the orthogonal direction of $\pm0.17\%$.  The O I
7774\AA\ is almost completely described by $P_{d}$, with nearly no
signal or variability in $P_{o}$ over the entire line profile
(Fig. \ref{obsres:dom:epoch2}).  At day 256, 2002 Aug 16, the line is
observed in emission, and there is no significant polarization
signature associated with it.
\subsubsection{Fe II}
\label{feiisec}
\begin{figure}
\figurenum{11c}
\hfil
\rotatebox{-90}{
\includegraphics[width=10cm]{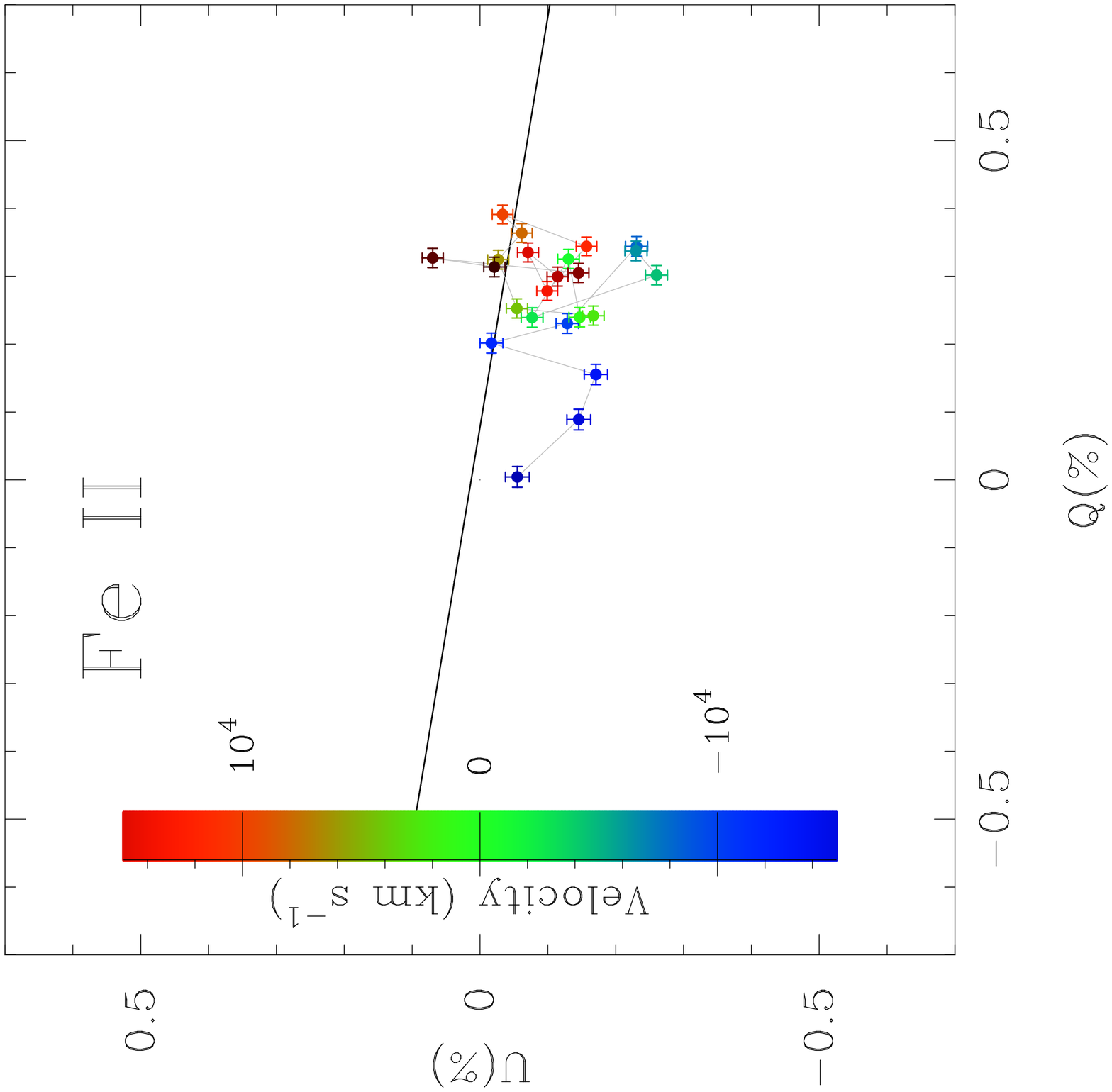}}
\hfil
\caption{}
\label{disc:quloop:epoch1:c}
\end{figure}

The wavelength range from 4800-5400\AA\ is expected to be mostly
composed of Fe II lines, with their overlap (due to Doppler
broadening) leading to depolarization across this wavelength range
\citep{2001ApJ...556..302H}.  At day 13, there is a net polarization
of $0.2\%$ observed across this range.  This is likely due to the
presence of strong polarization signatures from other lines, due to
other species, which cannot be completely depolarized by the Fe II
lines; for example, Fe II 4924\AA\ blended with H$\beta$\ and He
4921\AA, and 5018\AA\ blended with He 5015\AA\
\citep{2000AJ....120.1487M}.  The Stokes parameters show no preferred
orientation across the Fe II lines (Figs. \ref{obsres:dom:epoch1} and
\ref{disc:quloop:epoch1:c}) and a rise in the dominant rotated Stokes
parameter across the wavelength range is interpreted as being due to
blending with other species.  At day 31, a net depolarization from
0.5\% to 0.1\%, compared to other features, is observed across this
wavelength range.  The effects of possible blending are illustrated by
the proximity of H$\beta$, blueward, and He I 5876\AA\ redward.
Taking the wavelength range of 4800-5400\AA\ as being representative
of a number of Fe II lines, the data on the $Q-U$ plane and the
dominant and orthogonal rotated Stokes parameters shows that the
associated polarization is well described by the dominant axis (see
Figs. \ref{obsres:dom:epoch2} and \ref{disc:quloop:c}) with minor
deviations due to the behaviour specific to individual lines of either
Fe II or blends of Fe II lines with those other species
\citep{2000AJ....120.1487M}.  The cause of the polarization peak at
the second epoch at 4750\AA\ is unclear, given the high density of Fe
II lines and H$\beta$.  The polarization feature is significantly
broader than any of the individual absorption features, in that
wavelength range, observed in the flux spectrum.  This implies that it
is unlikely that a single line is producing it.  This polarization is
only observed along the dominant axis (Fig. \ref{obsres:dom:epoch2}
and \ref{disc:quloop:c}), unlike \halpha.  It is, therefore, unlikely to
be due solely to H$\beta$ and since it would be expected to have similar polarization properties as \halpha.  It is more likely to
have arisen from Fe II lines as it is oriented along the dominant
axis,similar to the polarization properties measured for the Fe II
lines at larger wavelengths.

\subsubsection{Ca II IR triplet}
\begin{figure}
\figurenum{11d}
\hfil
\rotatebox{-90}{
\includegraphics[width=10cm]{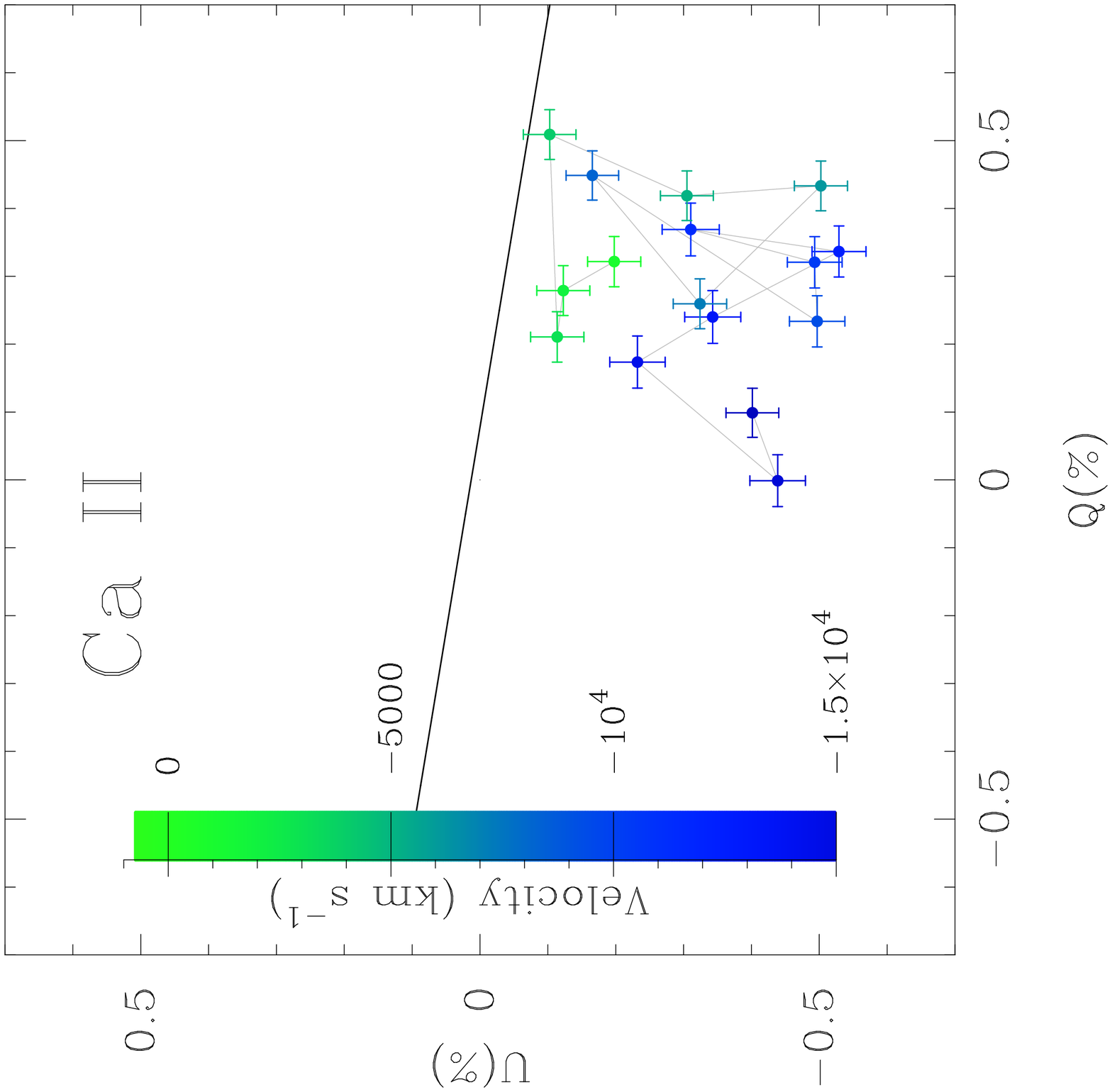}}
\hfil
\caption{}
\label{disc:quloop:epoch1:d}
\end{figure}

\begin{figure}
\figurenum{12a}
\hfil
\rotatebox{-90}{
\includegraphics[width=10cm]{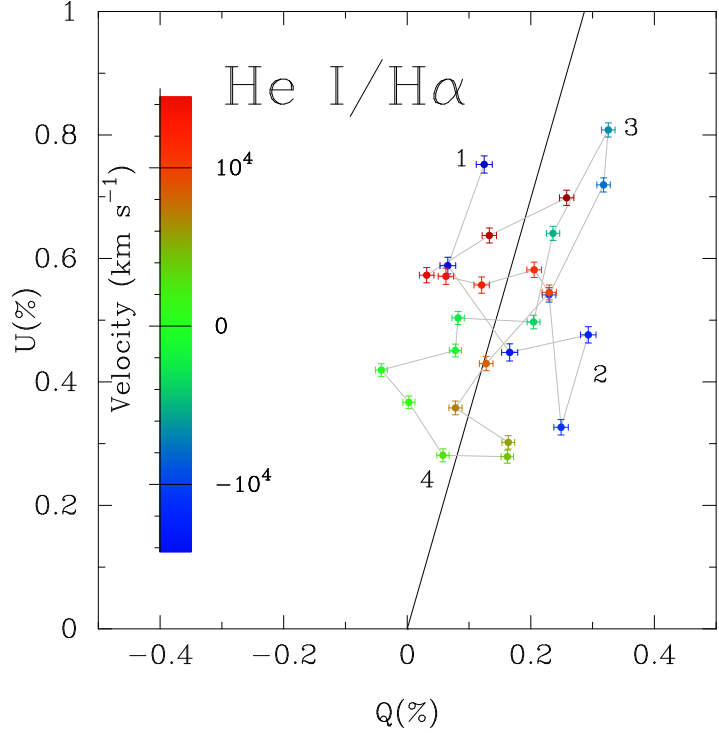}}
\hfil
\caption{Loops on the Q-U plane for a) \halpha/He I 6678\AA\ blend
(Points of interest are indicated by the numbers,
following the wavelength progression of the data on the Stokes plane:
1) Blue edge of the \halpha\ absorption profile in the flux spectrum;
2) Absorption minimum of \halpha, 3) Absorption minimum of He I
6678\AA; and 4) The \halpha\ and He I 6678\AA\ depolarizing emission
feature), b) O I 7774\AA, c) Fe II lines in the range 4800-5600\AA,\
and d) Ca II IR absorption from the observations of 2002 Jan 3.  The
data have been re-binned to 15\AA.  The heavy dashed line indicates
the dominant axis, calculated for the entire data set for this epoch
shown as shown on Fig. \ref{obsres:qu:epoch2}.  For Ca II the dominant
axis determined at the first epoch is shown as the dot-dashed line.}
\label{disc:quloop:a}
\end{figure}

\begin{figure}
\figurenum{12b}
\hfil
\rotatebox{-90}{
\includegraphics[width=10cm]{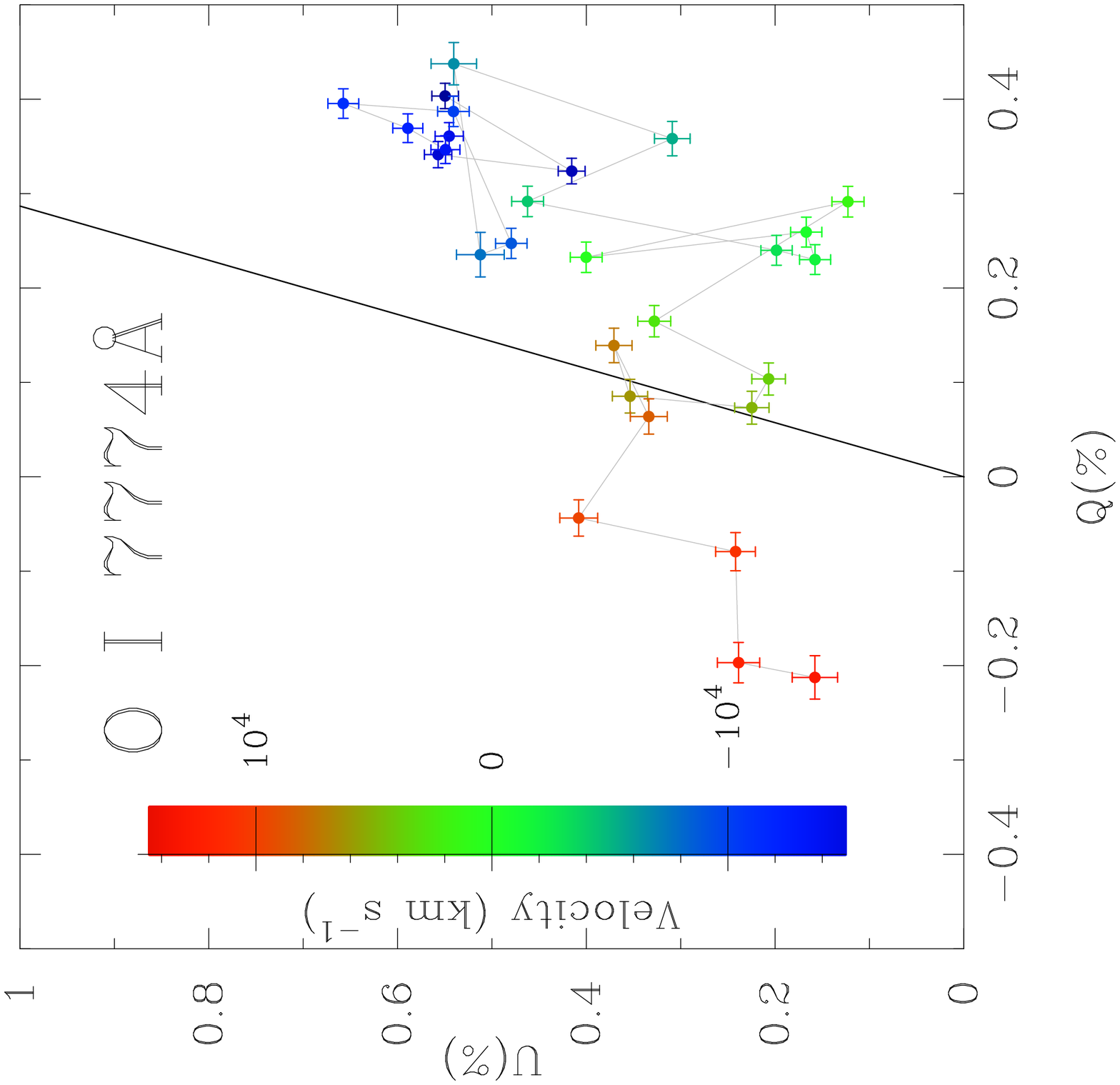}}
\hfil
\caption{}
\label{disc:quloop:b}
\end{figure}

\begin{figure}
\figurenum{12c}
\hfil
\rotatebox{-90}{
\includegraphics[width=10cm]{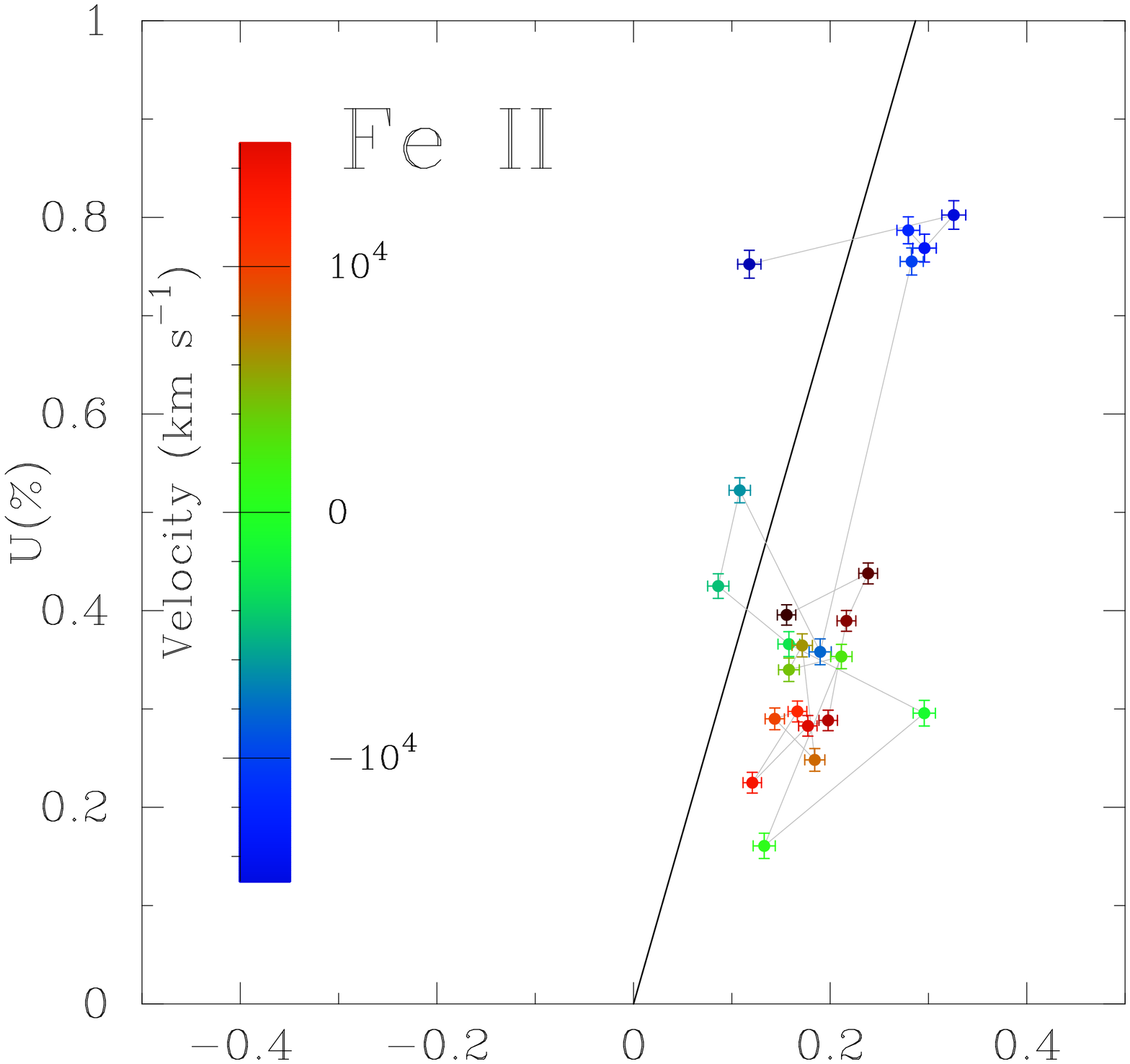}}
\hfil
\caption{}
\label{disc:quloop:c}
\end{figure}

\begin{figure}
\figurenum{12d}
\hfil
\rotatebox{-90}{
\includegraphics[width=10cm]{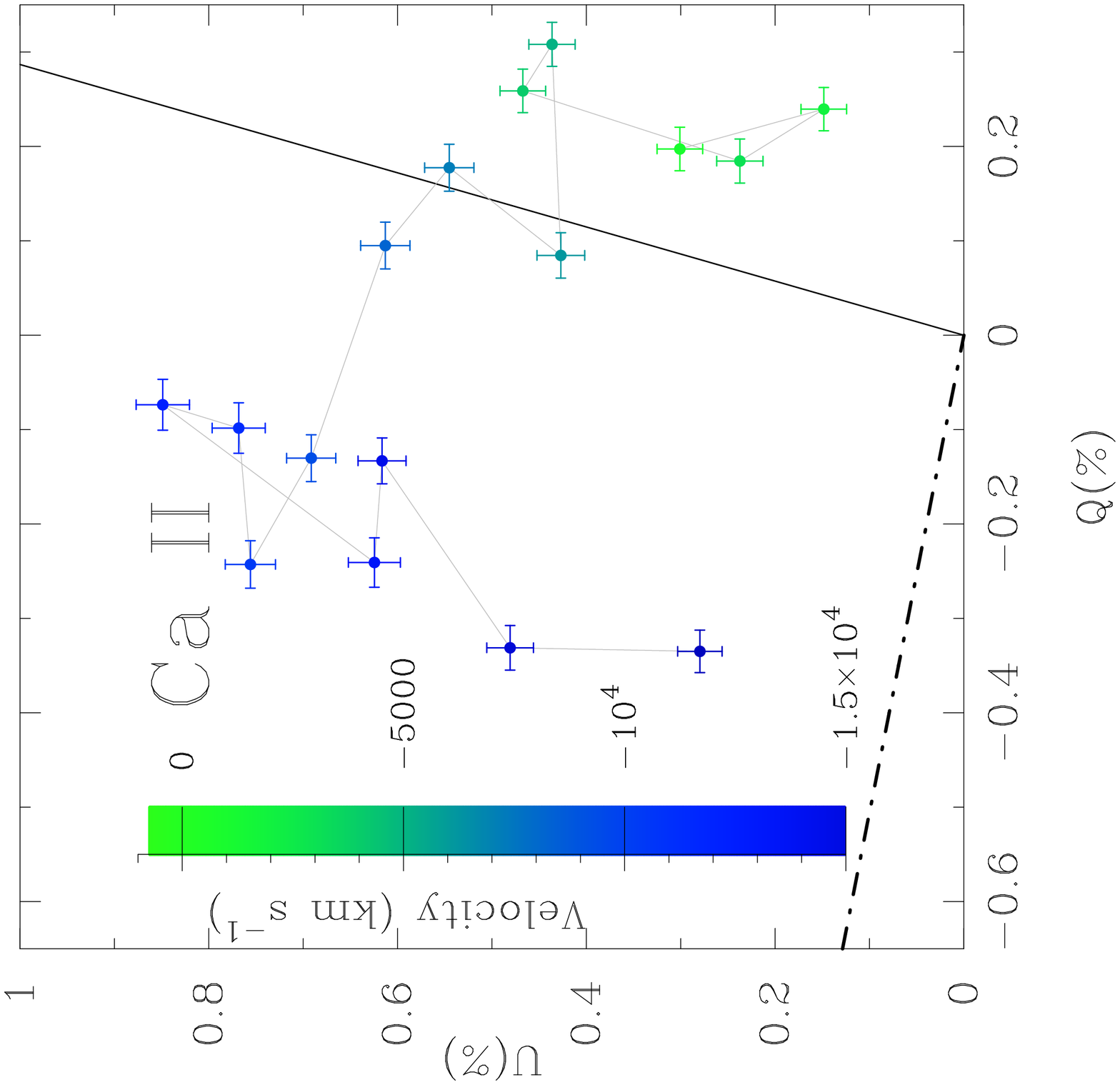}}
\hfil
\caption{}
\label{disc:quloop:d}
\end{figure}

The polarization properties of the Ca II IR triplet absorption feature
suffer from a number of effects which increase the levels of
uncertainties on the measurements.  The region redward of
$\mathrm{8000\AA}$ shows a significant drop in the response of the
FORS1 detector and hosts a particularly broad telluric absorption
band.  At the first epoch, there is a polarization associated with the
absorption feature, that reaches a maximum degree of polarization of
$0.4\pm0.3\%$ at the absorption minimum (see Fig. \ref{disc:quloop:epoch1:d}).  At day 31, with increased
S/N in the red, due to the increased brightness of the SN, the maximum
polarization is again observed to directly correlate with the
absorption minimum and is measured as $0.7\pm0.2\%$.  As for the O I
7774\AA\ line, the uncertainties of the measured Stokes parameters are
large and this is due to the telluric absorption band.  At the red
extreme, the profile of the Ca II IR blended triplet absorption, in
the flux spectrum, appears rising towards the emission component. The
polarization angle is observed to rotate from the continuum
polarization angle of $\sim40\degr$ through $+20\degr$ moving redward
towards the absorption minimum and back through $-20\degr$ when moving
from the absorption minimum toward the unobserved emission component.
On the $Q-U$ plane, see Fig. \ref{disc:quloop:d}, an incomplete loop is
observed, as the emission component was not observed.  The loop
appears to be aligned with the direction orthogonal to the dominant
axis (deviating by a maximum 0.4\%), and is approximately parallel to
the dominant axis determined for the observations of day 13.  The
rotated Stokes parameters $P_{d}$ and $P_{o}$ show that at day 13 the
polarization of the Ca II is parallel to the dominant axis
(Fig. \ref{obsres:dom:epoch1}).  By day 31, however, the polarization
has substantial components in both $P_{d}$ and $P_{o}$
(Fig. \ref{obsres:dom:epoch2}).  At day 256, there is no significant
polarization associated with the Ca II IR triplet, with large error
bars being symptomatic of the low S/N at these wavelengths.
\section{Discussion}
\label{disc}
The observations of SN 2001ig show the three distinct epochs in the
spectroscopic evolution of a Type IIb SN, accompanied with three
distinct sets of polarization properties.  The flux spectra of SN
2001ig at all three epochs are compared on Fig. \ref{obsres:flux} with
spectroscopic observations of SN 1993J, obtained from the SUSPECT
archive\footnote{http://bruford.nhn.ou.edu/$\sim$suspect/}.  At the
earliest epoch, SN 2001ig is observed to have a higher ratio of Balmer
to helium line strengths than SN 1993J.  Indeed, at this epoch the
spectrum of SN 2001ig is more similar to that of SN 1987K
\citep{1988AJ.....96.1941F,2002IAUC.7793....2C}, with the absorption
of He I 6678\AA\ at higher velocities and, therefore, being observed
to be blueward of \halpha.  The absorption minimum of \halpha\,
however, is observed to be significantly more blueshifted, by
$\mathrm{\sim 3 500}$\kms, than observed for 1987K.  Similarly, at 31
days the absorption of He I 6678\AA\ is at higher velocities than was
observed for SN 1993J at a similar epoch (12000 vs. 6940\kms;
\citealt{1993Natur.365..232S}).  In the nebular phase, at day 256, the
spectrum of SN 2001ig is very similar to SN 1993J at a similar age
showing [Ca II] 7291,7324\AA, He I 5876\AA\ and O I 6300/6364\AA.  The
ratios of the strengths of these lines in SN 2001ig are approximately
similar to those observed for SN 1987K and SN 1993J
\citep{1988AJ.....96.1941F,2000AJ....120.1487M,2006MNRAS.369L..32R},
although \citet{2002IAUC.7988....3F} note that the Mg I] 4471\AA\
observed for SN 2001ig is particularly strong (see
Fig. \ref{obsres:flux}).\\ At 13 days, when hydrogen features dominate
the spectrum, the general spectrum-wide polarization is low $\sim
0.2\%$.  This is consistent with the low polarizations observed for
Type IIP SNe \citep{2002Msngr.109...47W,2001ApJ...553..861L} which
have extended optically-thick hydrogen envelopes.  The presence of a
non-zero polarization implies that there is an asymmetry of the outer
hydrogen layers at the early times; which for an edge-on oblate
spheroidal configuration only requires an axis ratio of $\sim0.9$
\citep{1991A&A...246..481H}.\\ It is only at the second epoch when the
expected depolarized \halpha\ emission feature is first observed.
This is due to the decrease in the velocities of the ejecta material
compared to the first epoch, which changes the way in which the
absorption and emission features of \halpha\ and He I 6678\AA\ are
blended.  The increase in polarization of SN 2001ig between the first
and second epochs is also similar to other Type II SNe, when the
hydrogen layers cease to be optically thick and the interior
asymmetric He core is revealed (e.g. SN 1987A,
\citealt{1991ApJS...77..405J}; SN 2004dj,
\citealt{2006Natur.440..505L}).\\ The evolution in the polarization
signature of SN 2001ig at the first and second epochs is matched by
the spectroscopic evolution of the SN, with the increase in the
strength of helium lines by the second epoch.  The data on the $Q-U$
plane, at this epoch, are aligned with an obvious dominant axis which
suggests that the He core layer asymmetries still retain a strong
axial symmetry although clearly not a spherical symmetry
\citep{2003ApJ...592..457W}.  The dominant axis determined for the
data at the second epoch is, however, almost orthogonal to the
dominant axis of the data at the first epoch.  This suggests that
while asymmetries were present at day 13, they were not coupled to the
larger asymmetries present in the core that were revealed by day 31 in
either the degree of the asymmetry or the orientation of the asymmetry
on the sky.  This provides a case, therefore, that the hydrogen
envelope was sufficiently extended to not be directly coupled to the
asymmetries of the He core layers due to the explosion mechanism.
\citet{1995ApJ...440..821H} suggested that the small degree of early
polarization observed for SN 1993J was due to tidal distortion of the
progenitor by a binary companion, if the SN arose in such a system.
Alternatively, or in addition, another source of polarization at the
earlier epoch might be the aspherical distribution of sources of
excitation, from newly synthesized material, causing the shape of the
photosphere to deviate from a spherical symmetry, while the density
distribution of the hydrogen envelope remains physically spherically
symmetric \citep{2001AIPC..586..459H}.  The asymmetry measured at the
first epoch is, perhaps, an upper limit on the physical asymmetry of
the hydrogen envelope.\\ SN 2001ig and SN 1993J exhibit the same
evolution in polarization properties between days 13 and 31, with an
increase in the degree of the polarization accompanied by the rotation
of the polarization axis.  At the first epoch, the $V$-band
polarization of SN 2001ig (from synthetic broad-band polarimetry) is
0.3\% with $\theta=185\degr$ (ISP subtracted), whereas SN 1993J was
polarized at a level of $\sim 0.6\%$ at approximately the same epoch
(1993 Apr 7; \citealt{1997PASP..109..489T}).  At 31 days, the
broad-band $V$ polarization for SN 2001ig is $0.6\%$, whereas the
polarization of SN 1993J was $0.8\%$ (1993 Apr 26;
\citealt{1997PASP..109..489T}), and the polarization angle had rotated
through $\sim40\degr$.  In absorption components, however, SN 2001ig
is less polarized than SN 1993J by $\lesssim 1\%$.  The meandering
path of the V-band polarization of SN 1993J, as plotted on the $Q-U$
plane \citep[see Fig. 10 of][]{1997PASP..109..489T}, is also seen for
SN 2001ig; although the number of observational epochs of SN 2001ig is
much lower.\\ In these respects SN 2001ig is similar to 1993J, but
there are important exceptions.  At day 31 the polarizations of the He
I 5876\AA\ and 6678\AA\ absorption minima of SN 2001ig are comparable
to that of the \halpha\ absorption feature, whereas there is no
observed polarization signature attributable to He 6678\AA\ in SN
1993J.  This is due to the He I 6678\AA\ line for SN 2001ig being both
stronger and having a higher velocity than was observed for the same
line in SN 1993J.  The higher velocity for the absorption of He I
6678\AA\ means that it is not completely coincident with the
depolarizing \halpha\ emission flux, such that the polarization
signature of the absorption is still observed.\\
\citet{2001ApJ...550.1030W} suggested that the classification of a
CCSN as being of Type IIb might depend on the orientation of the SN as
seen by the observer, given the spectropolarimetric similarity between
SN 1993J and SN 1996cb.  The observations presented here have shown
that SN 2001ig arises from the same family of SNe as SN 1993J and SN
1996cb and, by extension due to spectroscopic similarity, to SN 1987K,
but that Type IIb SNe are not as homogeneous as suggested by
\citet{2001ApJ...550.1030W}.  In the context of Type I CCSNe, we note
that a whole range of polarizations have been measured for Type Ib/c
SNe, from $\sim1\%$ for SN 2002ap
\citep{2005Sci...308.1284M,2003ApJ...592..457W,2002PASP..114.1333L} to
$\gtrsim 4\%$ for the Type Ic SN 1997X (although this is a broad band
polarization measurement and it is unclear if this value is
representative of the continuum polarization;
\citealt{2001ApJ...550.1030W}).  On the other hand Type IIP and IIb
SNe exhibit a more limited range in polarization than the more
hydrogen deficient SNe.  This suggests that the retained hydrogen
envelope of the progenitor is important to dictating the polarization
evolution of Type IIb SNe, and that Type IIb SNe which have undergone
the transition to the hydrogen poor phase are not necessarily the same
as ``pure'' Type Ib SNe.\\ Particular note is made of the loop
structure observed, on the $Q-U$ plane, for certain spectral lines.
Loops were observed in spectropolarimetry of SN 1987A
\citep{1988MNRAS.231..695C} and similarly have been presented for the
Type Ia SN 2001el \citep{2003ApJ...593..788K}.  The production of a
loop requires an increase (or a decrease) in polarization across a
spectral line, accompanied by a rotation in polarization angle. While
\citet{2003ApJ...592..457W} discussed the ``dominant'' and
``orthogonal'' axes, a loop is a mixture of both.  The rotated Stokes
parameters of the Ca II IR triplet cannot be decomposed at day 31 to
two distinct orthogonal components (Fig. \ref{obsres:dom:epoch2}),
leading to loop structure on the $Q-U$ plane
(Fig. \ref{disc:quloop:a}).  The \halpha/He I blend is not, however, a
true loop since it can be decomposed into two orthogonal components
depending on which element dominates absorption at a particular
wavelength.  \citet{2003ApJ...593..788K} discussed how loops may be
formed by breaking a global axial symmetry by having varying degrees
of the departure of the ejecta from an axial symmetry as a function of
depth or with a clumpy SN ejecta.  The loops may hint at the presence
of plume-like structure in SN 2001ig ejecta, which is observed in SN
remnants \citep{1996ApJ...470..967F}.\\ \citet{2006ApJ...653..490W}
applied straight line fits to the Stokes parameters on the $Q-U$ plane
to judge the relative distributions of particular elements within the
ejecta of the Type Ia SN 2004dt.  If spectral lines produce loops
rather than a single dominant axis (i.e. a straight line) on the $Q-U$
plane, then straight-line fits of a dominant axis to a loop would be
inappropriate; this may then explain the high values of
$\chi^{2}_{\nu}$ that \citeauthor{2006ApJ...653..490W} found for
fitting dominant axes to a number of spectral features of SN 2004dt.\\
The locations of loops, hence the range of polarization angles, on the
$Q-U$ plane can be used in a similar manner to fits of the dominant
axes to particular spectral features to look for the relative
distributions of different elements within the SN ejecta.  For day 31,
species such as He and O are observed to have similar orientations on
the $Q-U$ plane and all lie along the dominant axis (see
Figs. \ref{obsres:dom:epoch2} and \ref{disc:quloop:c}), suggesting a
similar distribution for these elements in the ejecta.  The
polarization of \halpha, however, is predominantly orthogonal to these
species and, hence, has a significantly different distribution within
the ejecta.  The Stokes parameters across the Ca II absorption
feature, however, seem to have a similar orientation to the dominant
axis measured for day 13.\\ The nature of the progenitor system,
single or binary, that gave rise to SN 2001ig has been subject to much
debate.  \citet{2004MNRAS.349.1093R} observed periodic fluctuations in
the radio light curve of SN 2001ig, due to fluctuations in the CSM
density.  \citeauthor{2004MNRAS.349.1093R} concluded that this was due
to the presence of a binary companion, disturbing the stellar wind
from the progenitor as the immediate CSM was formed, with the
periodicity in the density fluctuations corresponding to the orbit of
the companion.  \citet{2006A&A...460L...5K} suggested, however, that
episodic mass loss from a Luminous Blue Variable phase of the
progenitor could also produce similar density fluctuations in the CSM.
Additionally \citet{2006ApJ...651.1005S} showed that SN 2003bg had a
similar radio light curve to SN 2001ig; the likelihood of two systems
having the same orientation and binary progenitor system parameters,
to produce such similar light curves, was considered to be too low to
make a binary progenitor a plausible scenario based on radio
measurements alone.  In the case of the Type IIb SN 1993J, the
progenitor was identified in fortuitous pre-explosion images
\citep{alder93j} and the companion was identified in very late time
images and spectroscopy \citep{maund93j}, demonstrating the production
of that particular Type IIb SN from a binary progenitor system.
\citet{2006MNRAS.369L..32R} have claimed to observe a possible stellar
source, 3 years post-explosion, at the site of the fading remnant of
SN 2001ig.  It is, therefore, particularly interesting to discuss the
observations presented here in the context of the nature of the
progenitor system.  \citet{2001ApJ...550.1030W} concluded that Type
IIb SNe, having so many similar spectropolarimetric properties, must
arise with similar orientations to the observer.  The shared
characteristics of SN 2001ig with Type IIb SNe 1993J and 1996cb, while
not a complete match, might suggest a common type of progenitor with
similar orientation; thus favouring a binary progenitor.  The
differences in the polarization properties observed for SN 2001ig days
13 and 31 suggests that the hydrogen and core layers did not possess
the same asymmetry, both in degree and orientation.  The hydrogen
envelope was, therefore, sufficiently large (in radius) to not be
subject to the same asymmetries as the core; this is in contrast to
the picture of \citet{2006ApJ...651.1005S}, who suggest a thin
hydrogen layer on a Wolf-Rayet progenitor.\\ In both the jet-torus
model \citep{2001AIPC..586..459H} and the binary progenitor system
model \citep{1995ApJ...440..821H}, it is clear how different
asymmetries of the envelope and the core-layers (defined by the
physically orthogonal equatorial plane and rotational axis of the progenitor)
would lead to different polarization angles for the dominant axis
between the first two epochs.  In the jet-torus model, the production
of Ni in a jet would cause asymmetric excitation of the hydrogen
envelope aligned with the rotational axis, while the core-layers would
form a torus in the equatorial plane.  In the binary progenitor model,
the asymmetries of the hydrogen envelope would be caused by tidal
interaction with the companion, causing deformation in the direction
of the equatorial/orbital plane, and a jet, that did not breach the
outer core-layers, would cause elongation of the core-layers along the
rotational axis.  Neither of these two simple models can explain the
physical rotation of the asymmetries, between the first and second
epochs, of only $\sim40\degr$ when both favor a rotation of
$90\degr$.\\ Alternatively, \citet{maund05bf} present a "tilted jet"
model.  In this model the jet axis is not aligned with the rotational
axis of the progenitor, such that the outer layers and the inner core
layers do not have to have either the same or orthogonal axes of
symmetry.  The outer layers, in this model, retain the shape of the
progenitor, but the core-layers are elongated by the mechanical force
of the jet and the redistribution of Ni along the jet axis.  In the
tilted-jet model, the angle between the axis of symmetry of the core
and the outer layers is not fixed.  We note that the polarization axis
of SN~1987A is tilted by $\sim 15\degr$ with respect to that of the
circumstellar rings \citep{2002ApJ...579..671W} and similar deviations
might characterize Cas A \citep{ourcasa}.  The Standing Accretion
Shock Instability in 3-D may provide a mechanism through which the
axis at which the jet is launched is no longer aligned with the
rotational axis of the progenitor, with \citet{mezza3d} modelling the
rotational axis of the proto-neutron star, in 3-D core-collapse
simulations, becoming misaligned by angles such as $10\degr$,
$15\degr$ and $45\degr$ from the rotational axis of the Fe core.
Other non-axisymmetric instabilities may also play a role
\citep{2007ApJ...654..429W}.\\ It is clear, however, that along with
appropriate models spectropolarimetry data can provide important
insight into the nature of the explosion mechanism and the progenitors
of this type of CCSNe.
\section{Conclusions}
\label{conc}
The Type IIb SN 2001ig has been observed to have a strong polarization
signature, consistent with other Type IIb SNe.  The polarization
properties are generally in excess of those observed for Type IIP SNe,
but at the lower limit of polarizations observed for the hydrogen
deficient Type Ibc SNe. The ISP is constrained by analyzing the
polarization behaviour of SN 2001ig at three separate epochs, but
particularly using observations at the last epoch, in the nebular
phase, when SN 2001ig is believed to be intrinsically unpolarized.  SN
2001ig has a low continuum polarization of 0.3\% at 13 days,
consistent with an almost spherical photosphere (with deviation of
$\lesssim10\%$) and the spectrum is dominated by hydrogen features.
At 31 days the decrease in the strength of the hydrogen lines and the
increase in the strength of the helium lines, along with a sharp
increase in continuum polarization to $\sim 1\%$, demonstrate that at
this epoch the highly asymmetric He core is being revealed.  The
absorption components of He I 5876\AA\ and 6678\AA\ are observed at 31
days to be highly polarized (to $\sim0.8\%$).  Loops, on the $Q-U$
plane, are observed for \halpha/He I 6678\AA, O I 7774\AA\ and the Ca
II IR triplet at 31 days.  The presence of these loop features
indicates a deviation of the SN ejecta from a global axial symmetry.
There is significant rotation of the observed Stokes $Q$ and $U$
parameters, between the first and second epochs, through an angle of
$\sim40\degr$; indicating that the hydrogen envelope was sufficiently
extended to not be coupled to the asymmetries of the He core-layers,
associated with the explosion mechanism.  The different geometries of
the outer envelope layers and the inner core region may be evidence of
the ``tilted jet'' model.  SN 2001ig is intrinsically unpolarized at
256 days.  The different polarizations at the three epochs show the
same physical process observed in Type IIP SNe, of an almost
spherically symmetric hydrogen layer shielding a highly asymmetric He
core.  As time since explosion increases the hydrogen layer becomes
optically thin revealing the He core and the polarization is seen to
increase.  The expansion velocities of the hydrogen and helium in SN
2001ig are significantly higher than observed for 93J-like Type IIb
SNe, but are similar to those observed for SN 1987K, and the
polarization properties of SN 2001ig show distinct differences from SN
1993J and SN 1996cb, which are mostly due different degrees of line
blending.  The differences between SN 2001ig and SNe 1993J and 1996cb
suggest that Type IIb SNe are not as homogeneous as perhaps previously
considered, but may be sufficiently homogeneous to share similar
geometries and, potentially, progenitor systems.
\section*{Acknowledgements}

The authors are grateful to the European Southern Observatory for the
generous allocation of observing time. They especially thank the staff
of the Paranal Observatory for their competent and never-tiring
support of this project in service mode.  The research of JRM and JCW
is supported in part by NSF grant AST-0406740 and NASA grant
NNG04GL00G.  \bibliographystyle{apj}

\newpage

\begin{table}
\caption{\label{obstab}Journal of Spectropolarimetric Observations of SN 2001ig}
\begin{tabular}{ccccccc}
\hline\hline
Star      & Date           & Exposure        & Wavelength  & Air Mass & Filter & Comment\\
          &                &  (s)            &($\rm{\AA}$) &          &        &        \\
\hline
SN 2001ig & 2001 Dec 16.11 & $4 \times 1000$ & 4450-8635   & 1.86     & GG435  &   SN   \\
HD 49798  & 2001 Dec 16.12 & $10$            & 4450-8635   & 1.31     & GG435  &Flux STD\\
LTT 9491  & 2002 Jan 03.02 & $200$           & 4450-8635   & 1.47     & GG435  &Flux STD\\
SN 2001ig & 2002 Jan 03.07 & $4 \times 500$  & 4450-8635   & 2.19     & GG435  &   SN   \\
EG 274    & 2002 Aug 15.98 & $60$            & 3700-8635   & 1.03     & None   &Flux STD\\
SN 2001ig & 2002 Aug 16.26 & $4 \times 1200$ & 3700-8635   & 1.04     & None   &   SN   \\
GD 50     & 2002 Aug 16.44 & $60$            & 3700-8635   & 1.09     & None   &Flux STD\\
\hline\hline
\end{tabular}
All observations conducted with the GRIS-300V grism.
\end{table}
\begin{table}
\caption{\label{linestab} Prominent P Cygni Profiles and Velocities at Absorption Minimum at +13 and +31 days}
\begin{tabular}{ccc}
\hline\hline
                 &\multicolumn{2}{c}{Epoch}      \\
                 &  $+13$ days    & $+31$ days   \\
Line             &                &              \\ 
\cline{2-3}
                 &  $v$\ (\kms)    & $v$\ (\kms)  \\
Fe II 5169\AA    &  blended        &  -10 300      \\
He I 5876\AA     &  -14 000        &  -12 000      \\
Si II 6355\AA    &   \ldots        &   -8 000      \\
\halpha          &  -19 000        &  -13 600      \\
He I 6678\AA     &  -19 000        &  -12 000      \\
O I 7774\AA      &  -12 700        &  -10 600      \\
Ca II IR triplet &  -12 700        &  -12 400      \\
\hline\hline
\end{tabular}
\end{table}

\clearpage
\clearpage
\clearpage
\clearpage
\clearpage
\clearpage
\clearpage
\clearpage
\clearpage
\clearpage

\clearpage
\clearpage
\clearpage
\clearpage
\clearpage
\clearpage

\clearpage

\clearpage

\end{document}